\documentclass[fleqn,usenatbib]{mnras}
\usepackage{newtxtext,newtxmath}
\usepackage[T1]{fontenc}
\DeclareRobustCommand{\VAN}[3]{#2}
\let\VANthebibliography\thebibliography
\def\thebibliography{\DeclareRobustCommand{\VAN}[3]{##3}\VANthebibliography}

\usepackage{breqn}
\usepackage{amsmath}
\usepackage{graphicx}
\usepackage{units}
\usepackage{sidecap}
\usepackage{float}
\sidecaptionvpos{figure}{c}
\usepackage{longtable}
\usepackage{multirow}
\usepackage{amsfonts}
\usepackage{array}
\usepackage{tabularx}
\usepackage{orcidlink}

\newcommand{\Msun}{${\rm M}_{\odot}$}
\newcommand{\Mcrit}{${M_{\rm crit}}$}
\newcommand{\JLW}{${J_{\rm LW}}$}
\newcommand{\Htwo}{$\rm {H_{2}}$}

\newcommand{\vbc}{$v_{\rm bc}$}

\newcommand{\tdelay}{$t_{\rm delay}$}
\newcommand{\overdense}{$\delta(\vec{x})$}

\newcommand{\TS}{$T_{\mathrm{S}}$}
\newcommand{\TK}{$T_{\mathrm{K}}$}
\newcommand{\TC}{$T_{\mathrm{C}}$}
\newcommand{\Tb}{$\delta T_{\mathrm{b}}$}
\newcommand{\Tbmin}{$\delta T_{\mathrm{b,min}}$}

\newcommand{\xe}{$x_{\mathrm{e}}$}
\newcommand{\xa}{$x_{\alpha}$}
\newcommand{\xc}{$x_{\mathrm{c}}$}
\newcommand{\Lya}{Ly-$\alpha$}
\newcommand{\Ja}{$J_{\alpha}$}
\newcommand{\JX}{$J_{\mathrm{X}}$}

\newcommand{\SFRDII}{$\mathrm{SFRD_{II}}$}
\newcommand{\SFRDIII}{$\mathrm{SFRD_{III}}$}
\newcommand{\PS}{$\Delta^2(k)$}

\defcitealias{Paper1}{F24}
\defcitealias{Paper2}{F25}
\defcitealias{Hazlett25a}{H25a}
\defcitealias{Hazlett25b}{H25b}
\defcitealias{Munoz22}{M22}
\defcitealias{Hirata06}{H06}

\title[Simulating the High-Redshift 21-cm Signal]{Self-Consistent Simulations of the High-Redshift 21-cm Signal with Star Formation Calibrated to Hydrodynamic Simulations}

\author[C. R. Feathers et al.]{
Colton R. Feathers,$^{1,2}$\thanks{E-mail: 31022251@sun.ac.za}\orcidlink{0000-0001-9062-4615}
Eli Visbal,$^{1}$\thanks{E-mail: elijah.visbal@utoledo.edu}\orcidlink{0000-0002-8365-0337}
Steven G. Murray,$^{2}$\thanks{E-mail: sgmurray@sun.ac.za}\orcidlink{0000-0003-3059-3823}
Ryan Hazlett,$^{1}$\thanks{E-mail: ryan.hazlett@rockets.utoledo.edu}\orcidlink{0000-0002-1034-7986}
and Yin-Zhe Ma,$^{2}$\thanks{E-mail: mayinzhe@sun.ac.za}\orcidlink{0000-0001-8108-0986}
\\
$^{1}$Ritter Astrophysical Research Center, University of Toledo, 2855 W. Bancroft St., Toledo, Ohio 43606, USA\\
$^{2}$Department of Physics, Stellenbosch University, 111 Merriman Ave, Stellenbosch, Western Cape 7600, South Africa\\
}

\date{Accepted XXX. Received YYY; in original form ZZZ}
\pubyear{\the\year{}}
\begin{document}
\label{firstpage}
\pagerange{\pageref{firstpage}--\pageref{lastpage}}
\maketitle


\begin{abstract}
We present a novel, self-consistent, semi-numeric Cosmic Dawn (CD) simulation in which small-scale star formation (SF) is calibrated to the \emph{AEOS} and \emph{Renaissance} hydrodynamic simulations. SF proceeds within dark matter (DM) halos via neural network emulation while considering large-scale fluctuations in density and feedback. We translate the resulting 3D distribution of galaxies into predictions for the 21-cm brightness temperature, \Tb, and power spectrum, \PS. We simulate several unique realizations to study the impact of varying astrophysics on \Tb, finding that more efficient Population II (PopII) SF largely yields stronger Lyman-$\alpha$ coupling, resulting in a shallower and wider absorption trough. However, we find that PopII SF dominates \PS\ at $z \lesssim 20$ and on smaller scales at intermediate redshifts ($k \gtrsim 0.2\ \mathrm{Mpc^{-1}}$ at $z \simeq 34-20$) while Population III (PopIII) SF dominates \PS\ at $z\gtrsim34$ and on larger scales at intermediate redshifts. Compared with previous works, we find that the combination of hydrodynamic SF calibration, a critical halo mass for SF considering \Htwo\ self-shielding, and stochastic DM halo merger histories results in both earlier SF and higher SF rates across CD. Further, we find that the delay period separating PopIII and PopII SF (\tdelay) significantly impacts \Tb, and that one must include DM halo merger histories to properly account for this transition. Finally, we find our fiducial \Tb\ to be detectable at $z\lesssim25$ with 1080 hours of HERA observations under moderate foreground assumptions, and the lack of such a detection at $z \gtrsim 20$ would suggest \tdelay\ $\gtrsim$ 30 Myr.
\end{abstract}



\section{Introduction} \label{Intro}
The first stars and galaxies are expected to have formed within the first few hundred Myr years of cosmic history, during an era referred to as Cosmic Dawn (CD). The astrophysical details of CD are crucial for our understanding of modern cosmic evolution, and so significant efforts have been made to advance our understanding of how the universe developed from a largely uniform volume of primordial gas, radiation, and dark matter into the highly clustered, ionized cosmos we observe locally \citep[e.g. see][for reviews]{Sitwell2014, Somerville15, Greif15, Wise19, Inayoshi20-BH_Review, Klessen-Glover23, Zhang2023, Zhou2025, Cang2025, Zhang2026}. 

Observational probes of CD may be obtained through state-of-the-art observatories such as \emph{JWST} with its powerful infrared sensitivities which allow for high-redshift observations of early star and galaxy formation \citep[e.g.][]{Finkelstein23, Matthee24, Naidu25}, though even \emph{JWST} is limited to the brightest sources in the early stages of CD. On the other hand, large-scale radio observatories such as HERA \citep[e.g.][]{DeBoer17, HERA-23}, LOFAR \citep[e.g.][]{vanHaarlem13-LOFAR, Gehlot19}, and MWA \citep[e.g.][]{EwallWice16, Yoshiura21} may study the evolution of the cosmos through wide-field observations at various cosmological redshifts, yielding a deeper, more large-scale statistical view of CD \citep[also see][]{Mellema13, Price18, Eastwood19, Gehlot20, Garsden21, Munshi25}. 

Interpreting such observations then requires the development of theoretical models that predict the temporal and spatial evolution of both the earliest luminous sources, and their impact on the broader intergalactic medium (IGM) \citep[e.g.][]{Kulkarni19, Vikaeus22, Garaldi24, Trinca24}. Such models are statistically compared to observational data in order to yield constraints on physical mechanisms of CD and aid our understanding of cosmic evolution \citep[e.g.][]{Armengaud17, Kehrig18-HeII, Mauerhofer25, Visbal25}.

One of the more sought after observational constraints for CD is that of the 21-cm line \citep[see][for reviews of 21-cm physics]{Furlanetto06, Pritchard12, Liu-Shaw20}. Because the primordial gas of the early universe was largely composed of hydrogen and helium, this relatively rare spin-flip transition of neutral atomic hydrogen may be seen shining across CD. Then, as the first astrophysical sources formed and began to emit ionizing radiation, the 21-cm signal weakened as evermore of the early IGM became ionized with time. This renders 21-cm observations as a powerful probe into cosmic evolution as it directly traces the ionization history of the cosmos \citep[e.g.][]{Cohen17, Ahn21, Ventura23, Ventura25, Sikder24, Zhang24-21cm}. By observing the evolution of the globally averaged 21-cm signal and its power spectrum (i.e. the spatial fluctuations in the signal), one may then place significant constraints on the rise of the first stars and galaxies, their properties, and the processes that dominated cosmic evolution prior to the end of the Epoch of Reionization (EoR, $z \sim 6$) \citep[e.g.][]{Dayal20, Liu-Shaw20, Kaur2022, Barry22, CC-26}. 

Recognizing the potential that such observations may hold for future astrophysical research, there has been a significant rise in the number of works simulating the cosmological 21-cm signal in recent years \citep[e.g.][]{Fialkov14a, Fialkov14b, Clark2018, Magg22, Munoz19b, Munoz23, Mittal24, Cruz24, Cruz25, Liu25}. Such simulations have assumed many different forms throughout the literature as the properties of the first stars, their formation, and contribution to cosmic reionization remain largely unconstrained \citep[][]{Klessen-Glover23}. This lack of observational constraints forces simulators to make simplifying assumptions in their theoretical frameworks, resulting in a wide variety of simulation techniques, each with unique strengths and weaknesses for providing 21-cm signal predictions \citep[see][for reviews]{Greif15, Somerville15, Moriwaki23}. 

Predictions for the 21-cm signal rely on some underlying model for CD and the rise of the first stars and galaxies. Contemporary CD/EoR models range from simplified analytic estimates \citep[e.g.][]{Visbal15a, Hartwig18a, Furlanetto22, Munoz22, Munoz23, Cruz24, Liu24} to complex (magneto-)hydrodynamic simulations that account for as much astrophysical detail as possible \citep[e.g.][]{Xu13-Renaissance, Xu14-Renaissance, Bryan14-ENZO, OShea15-Renaissance, Xu15b-Renaissance, Xu16-Renaissance, Barrow17, Norman18, Jaacks18, Garaldi24}. While global analytic models are efficient and may be used to explore large parameter spaces, they are naturally limited by their simplifying assumptions. Conversely, complex hydrodynamic simulations necessitate large computation times and resources which limit them to relatively small simulation volumes and narrow parameter spaces, limiting research to highly specific areas of interest while the interplay between various parameters is largely ignored.

Here we implement an intermediate, semi-numeric simulation framework that is calibrated to hydrodynamical simulations, including 3D spatial information, a self-consistent treatment for radiative feedback, and stochastic dark matter (DM) halo merger histories. Such considerations yield more accurate results than analytic models while also maintaining the flexibility to allow for more realizations than hydrodynamic simulations. This framework may be viewed as a novel, complimentary simulation to those of 21cmFAST \citep[e.g.][]{Mesinger11-21cmFAST, Davies25}, Zeus21 \citep{Cruz24, Cruz25}, and other similar works aiming to provide semi-numeric 21-cm predictions \citep[e.g.][]{Fialkov14a, Fialkov14b, Cohen17, Fialkov19, Sikder24, Jones25}. The results of this work may then be compared with those of previous models to quantify how different techniques and assumptions impact the predicted 21-cm signal while providing theoretical constraints for the expected signal which future observational experiments may then verify or rule out.

We aim to provide accurate theoretical predictions of the 21-cm global signal, power spectrum, and their evolutions across CD by extending the semi-numeric simulation framework presented in \cite{Paper2}, which provided self-consistent, large-scale predictions of the star formation rate density (SFRD) from $z = 60-15$ through the use of artificial neural network emulation. This framework was the first to self-consistently unite small-scale DM halo merger trees containing minihalos with their large-scale Lyman-Werner feedback from early star formation (SF). We improve upon this framework by increasing its computational efficiency, including SF prescriptions calibrated to state-of-the-art cosmological hydrodynamic simulations, and developing a post-processing framework that translates the SFRD results of the simulation volume into novel predictions for the cosmological 21-cm signal. Improving the efficiency of our framework also allows us to compare our fiducial results with realizations which assume different SF models to study how each impacts the large-scale results.

The most significant improvement made to the simulation framework of \cite{Paper2} (henceforth \citetalias{Paper2}) is our upgraded subgrid SF prescription; here we adopt the model presented in \cite{Hazlett25b}, which calibrates Population II and III (PopII and PopIII) SF to the results of hydrodynamical simulations. We then compare the results of this upgraded fiducial framework to those resulting from the framework of \citetalias{Paper2} to show their impact. We also compare our results with models that vary the PopIII stellar mass introduced at each SF event, models which delay the onset of PopIII and/or PopII SF, and also a simplified analytic SF model similar to those presented in \cite{Mashian16, Munoz23}. Each distinct realization models the subgrid SF history for our simulation volume with differing approaches, and we aim to study the impacts that each have on the 21-cm global signal, its power spectrum, and their evolution across CD.

The rest of this paper is organized as follows. In \S \ref{Methods} we review the simulation framework of \citetalias{Paper2}, outlining the changes and improvements made for the fiducial model of this work. We also outline the neural net training process and the necessary calculations for translating our SF results into estimates for the 21-cm signal. In \S \ref{Fid_Results} we discuss the results of our fiducial simulation. We then compare our fiducial results with our alternative simulations as well as previously reported values from the literature, and provide observability estimates with HERA in \S \ref{Results_Compare}. Finally, we summarize our results in \S \ref{Conclusions} and discuss planned future work with this new 21-cm simulation framework. Unless otherwise stated, all distance scales are in comoving units, all halo masses and cosmic overdensity values are those at $z=15$, and all baryon-DM streaming velocities are those found at Recombination. Also note that all Lyman-Werner background intensity values referred to in this work have units of $J_{\rm 21} = 10^{-21}\ \rm erg\ s^{-1}\ cm^{-2}\ Hz^{-1}\ sr^{-1}$. Finally, we assume a $\Lambda$CDM cosmology throughout, consistent with \cite{PlanckCollaboration}, adopting the following parameter values: $\Omega_{\rm m} = 0.32$, $\Omega_{\rm \Lambda} = 0.68$, $\Omega_{\rm b} = 0.049$, $h = 0.67$, and $\sigma_8 = 0.83$.

\section{Simulation Methods} \label{Methods}
In this section, we outline the details of our simulation framework and the relevant astrophysics necessary for estimating the 21-cm signal. In \S \ref{sub:SF_Model},  we describe the methods for generating our large-scale simulation volume and the training process for our subgrid SF emulators, outlining any modifications/extensions made to the framework of \citetalias{Paper2}. We then specify the fiducial model for this work, as well as several alternative subgrid models to which we compare our fiducial results. Then, in \S \ref{sub:21cm_Model}, we present a detailed calculation for translating our simulated SF results into predictions for the 21-cm global signal and power spectrum. 

\subsection{Semi-numeric Star Formation Framework} \label{sub:SF_Model}

\subsubsection{Simulation Volume} \label{subsub:Sim_Volume}
In order to accurately simulate SF across a large cosmic volume, we must include spatial fluctuations in matter density, which further necessitates modelling the relative baryon-DM streaming velocities resulting from those fluctuations \citep{Tseliakhovich10-vbc, Tseliakhovich11}. Such spatial information allows for a more accurate SFRD predictions and thus self-consistent treatment for radiative feedback than simulations which assume a constant matter density and/or streaming velocity field, and is also crucial for yielding accurate power spectra.

The simulation volume used here is identical to that of \citetalias{Paper2} --- 192 Mpc on a side, consisting of $64^3$ $(3\ \mathrm{Mpc})^3$ cells, with each cell assumed to have a constant cosmic overdensity, \overdense, and baryon-DM streaming velocity, \vbc\ (see referenced Fig. 2 for further detail). This simulation volume was generated using 21cmvFAST \citep{Munoz19b}, which included prescriptions for generating the \vbc\ field along with the underlying density field.

\subsubsection{Dark Matter Halo Evolution} \label{subsub:DM_Halos}
The backbone for many CD simulations is the underlying DM halo merger history, which describes the mass evolution of the halos which host the first stars and galaxies. Merger histories are typically constructed for a given simulation volume beforehand, then SF is iteratively modelled on top of said merger history as the simulation steps forward through time. 

In this work, DM halo evolution is modelled via Extended Press-Schechter (EPS) formalism \cite{Press74, Bond91-EPSformalism}. We construct Monte Carlo DM halo merger trees starting with the final halo mass at $z=15$, then step backward through time assuming binary mergers of halos separated in redshift space by $\Delta z = 0.05$ \citep[following a similar approach to][]{Lacey93}. As in \citetalias{Paper2}, we have generated a library of nearly $10^4$ DM halo merger trees for 40 halo mass bins, logarithmically spaced from $M_{\mathrm{halo}}(z=15) = 10^{5.6}-10^{9.5}$ \Msun. 

Merger histories are then constructed for our $(3\ \mathrm{Mpc})^3$ simulation cells according to their overdensity, effectively representing our spatial resolution scale. For each halo mass bin, abundances are Poisson sampled from an overdensity-dependent (`conditional') halo mass function \citep[CHMF, see Eqn.~(7) of][]{Barkana-Loeb04}, and the corresponding number of merger trees are randomly selected from the full library to construct the cell merger history. \footnote{As in \citetalias{Paper2}, we limit the maximum number of merger trees per halo mass bin to 100 for computational efficiency, finding that the average SF does not meaningfully change with the inclusion of more trees. In this case, we weight the average SF of that halo mass bin by the ratio of 100 to the Poisson sampled value to account for the truncation in halo abundance.}

We consider 400 overdensity bins spaced such that the total number of simulation cells within each bin is roughly constant \citepalias[see \S 2 of][for a more in-depth description of our DM halo merger history generation and binning]{Paper2}. Each overdensity bin therefore has its own unique DM halo merger history, which are assigned to the individual subgrid cells making up our simulation volume based on their \overdense, and serve as the scaffolding onto which we simulate SF\footnote{We note that this procedure artificially reduces the stochasticity of our DM halo merger histories by limiting the volume to 400 unique merger histories; however, it was found in \citetalias{Paper2} that important quantities such as the SFRD and power spectra are well converged with our chosen binning.}. 

\subsubsection{Star Formation Models} \label{subsub:SF_Model}
We now describe our fiducial subgrid SF model, as well as several comparison models used to generate the training data for our neural network emulators. We begin by reviewing the subgrid semi-analytic model (SAM) used in \citetalias{Paper2}, then discuss our upgraded fiducial model as well as a few alternative models with which we compare our results. Throughout this subsection, refer to Table \ref{Table1} for details of each subgrid SF model. 

The implementation of artificial neural network (NN) emulators was the focus of \citetalias{Paper2}, which served as a proof-of-principle that such NNs may be used to accurately simulate the subgrid SF of a more complex SAM. Their implementation was necessary because the SAM of \cite{Paper1} (henceforth \citetalias{Paper1}) was too inefficient to serve as the subgrid model for all $64^3$ cells of our large-scale simulation volume.\footnote{Even though the SAM of \citetalias{Paper1} may be realized in $\sim$200 seconds, the sheer number of halos that the SAM must cycle through and record properties for renders this computationally prohibitive for the full $(192\ \mathrm{Mpc})^3$ simulation volume, as each of the $64^3$ cells would require $\sim$200 seconds (amounting to $\sim14,500$ CPU-hours) and record the properties of all halos within them, drastically driving up memory requirements. The NN emulation framework of \citetalias{Paper2}, on the other hand, may be realized in $\sim$96 CPU-hours.} We therefore adopt this SAM as the training model whose results are then used to train the subgrid NNs of our simulation.

\begin{table*}
    \centering
    \begin{tabular}{|c|p{0.6\textwidth}|p{0.2\textwidth}|}
        \hline
         Model Name & Model Description & References \\ \hline \hline
         Fiducial & PopII and PopIII SF prescriptions calibrated to \emph{AEOS} and \emph{Renaissance} hydrodynamical simulations, including two-phase ``bursty'' PopII SF & \citetalias{Hazlett25a, Hazlett25b} \\
         \hline
         F25-Fid & Fiducial model of \citetalias{Paper2}, with minor efficiency improvements & \citetalias{Paper2} \\
         \hline
         Fid-100 & Same as ``Fiducial'' but introduces a single 100 \Msun\ PopIII star when halos first overcome \Mcrit\ & \citetalias{Paper2} \\
         \hline
         Fid-200 & Same as ``Fiducial'' but introduces a single 200 \Msun\ PopIII star when halos first overcome \Mcrit\ & \citetalias{Hazlett25a}; \citetalias{Paper1}: \citetalias{Paper2} \\
         \hline
         Fid-400 & Same as ``Fiducial'' but introduces a single 400 \Msun\ PopIII star when halos first overcome \Mcrit\ & \citetalias{Paper2} \\
         \hline
         Hi-\Mcrit\ & Same as ``Fiducial'' but raises \Mcrit($z$) by a factor of 2.25 & \citetalias{Hazlett25b} \\ 
         \hline
         Delayed & Same as ``Fiducial'' but with \tdelay\ = 100 Myr & e.g. \cite{Magg18, Magg22, GesseyJones22}, \citetalias{Hazlett25b} \\ 
         \hline
         HMF Integral & No NN emulation or DM halo merger histories. Integrates the HMF to determine the fraction of gas collapsed within halos which may then be used to estimate SF & e.g. \cite{Visbal15a, Mashian16, Munoz23}; \citetalias{Paper1} \\
         \hline
    \end{tabular}
    \caption{The various subgrid SF models used throughout this work (left), a brief description of their unique features (middle), and the corresponding literature references (right). Each provides a physically motivated framework for introducing stellar mass within DM halos, yielding distinct 21-cm signal predictions which we may then compare to assess the impact of each assumption.} \label{Table1}
\end{table*}

The SAM of \citetalias{Paper1} was designed to yield the globally-averaged SFRD for a given set of astrophysical parameters, and was the first such SAM to be calibrated to hydrodynamic simulations while considering self-consistent radiative feedback and the effects of \Htwo\ self-shielding within DM halo merger trees. This SAM was then transformed in \citetalias{Paper2} to operate for a finite volume of $(3\ \mathrm{Mpc})^3$, as this was found to be the smallest finite volume to yield convergent SFRD predictions with the global SAM (see Fig. 8 of \citetalias{Paper1}).

Stepping forward in time, the training SAM cycles through all DM halos at each redshift step and allows PopIII SF to proceed within pristine halos that are above the critical mass threshold, \Mcrit($z$). This threshold represents the minimum halo mass that allows for efficient radiative gas cooling, be it via molecular \Htwo\ or atomic hydrogen transitions. \Mcrit\ therefore depends heavily on the strength of the Lyman-Werner (LW) radiative background intensity, \JLW, and the \vbc. The former may dissociate \Htwo\ molecules and suppress halo gas cooling, while the latter suppresses primordial SF as the velocity of baryonic matter relative to the underlying DM halos sees gas advect out of low-mass halos with shallower gravitational potential wells. We also consider the effects of \Htwo\ self-shielding in our fiducial \Mcrit\ calculation, which is adopted from \cite{Kulkarni21} and is calibrated to the results of hydrodynamic simulations (see \S 3.2.1 of \citetalias{Paper1} for more detail).

Halos that have undergone PopIII SF in our SAM will then transition to enriched PopII SF following a constant delay period of $t_{\mathrm{delay}} = $ 10 Myr. This period represents the time it takes for PopIII stars to live, die, disrupt their host halo, and for the halo gas to recollect and cool once more. While assuming a constant delay period for all halos is not necessarily the most precise treatment for this transition, the exact threshold a halo must overcome to begin PopII SF is still largely uncertain. Some contemporary models assume a halo metallicity threshold, following the complexities of internal/external metal enrichment from PopIII supernovae \citep[e.g.][]{Greif15, Mebane18, Mebane20, Ventura23, Ventura24}, while others limit PopIII SF to molecular-cooling halos (MCHs, cooling via \Htwo) and PopII SF to atomic cooling halos \citep[ACHs, with $T_{\mathrm{vir}} \geq 10^4\ $K, e.g.][]{Munoz22, Cruz24}. There are even simulations where SF is not modelled as two distinct populations and are instead combined into one SF model \citep[e.g.][]{Mashian16, Park19}.

In the fiducial simulation of \citetalias{Paper2}, PopIII SF proceeds instantaneously, introducing a single 200 \Msun\ star once a halo overcomes \Mcrit. PopII SF then proceeds after \tdelay\ has elapsed, and follows the bathtub model presented in \cite{Furlanetto22} (see \citetalias{Paper2} for further detail). In this work, we denote this simulation framework as the ``F25-Fid'' model and compare its results to our new fiducial model described below.

For our updated fiducial simulation, we adopt the PopII and PopIII SF models of \cite{Hazlett25a, Hazlett25b} (henceforth, \citetalias{Hazlett25a} and \citetalias{Hazlett25b}, respectively) into our NN training data SAM. This represents the largest change to the simulation framework from \citetalias{Paper2}, and was implemented to yield more accurate SF results that are calibrated to the state-of-the-art hydrodynamical simulations of \emph{AEOS} \citep{Brauer25-AEOS} and \emph{Renaissance} \citep{OShea15-Renaissance}. The works of \citetalias{Hazlett25a} and \citetalias{Hazlett25b} present statistical fits for the PopII and PopIII stellar mass that will form within a given halo based on its mass. Here, we have incorporated these fits into our fiducial subgrid SAM, and by comparing with the F25-Fid results we may study how this calibrated SF model affects the resulting SFRDs and 21-cm signal.

The results of \citetalias{Hazlett25a} suggest an early ``bursty'' phase of PopII SF, followed by more ``steady'' SF once the host halo reaches $T_{\mathrm{vir}} \geq 1.76\times10^4$ K $\equiv T_{\mathrm{vir,steady}}$. During the bursty phase, halos undergo a burst in PopII SF then experience a quiescent period of no SF afterwards. This may occur multiple times within a single halo before it reaches $T_{\mathrm{vir,steady}}$, after which PopII SF proceeds following a similar prescription to our F25-Fid model.

More precisely, we implement the PopII SF models described by Eqns.~(1) \& (2) of \citetalias{Hazlett25a}, and Eqn.~(6) of \citetalias{Hazlett25b}. The former two equations are calibrated to the PopII SF of \emph{Renaissance}, allowing one to respectively calculate the stellar mass produced in a halo and its subsequent quiescent period. The latter equation is calibrated to \emph{AEOS}, which extends the \emph{Renaissance} SF equation to lower halo masses and higher redshifts. Ultimately, the PopII stellar mass produced in a halo during its bursty phase is given by
\begin{equation} \label{EQ:Bursty_II}
    M_{\mathrm{burst}} = \alpha {\left( \frac{M_{\mathrm{vir}}}{10^7\ {\rm M}_{\odot}} \right)}^\beta {\left( \frac{1+z}{20} \right)}^\gamma M_{\odot}, 
\end{equation}
where $\alpha$, $\beta$, and $\gamma$ are fitting parameters that depend on the mass of the halo. For those with $M_{\mathrm{halo}} < 4\times10^6$ \Msun, [$\alpha$, $\beta$, $\gamma$] = [6456.54, 0.4972, 0], adhering to the \emph{AEOS} calibration. For more massive halos ($M_{\mathrm{halo}} \geq 4\times10^6$ \Msun), [$\alpha$, $\beta$, $\gamma$] = [5364, 1.092, 0.578], as per the \emph{Renaissance} calibration. Following a burst in PopII SF, the host halo then enters a quiescent period during which it forms no stars, given by
\begin{equation} \label{EQ:t_quies}
    t_{\mathrm{quies}} = 44.81 \times {\left( \frac{M_{\mathrm{vir}}}{10^7\ {\rm M}_{\odot}} \right)}^{-0.394} \mathrm{Myr},
\end{equation}
regardless of whether the halo is in the \emph{AEOS} or \emph{Renaissance} mass range. After $t_{\mathrm{quies}}$ has elapsed, the halo may then undergo another burst in PopII SF following Eqn. \ref{EQ:Bursty_II}, or if it has grown to a large enough mass such that $T_{\mathrm{vir,halo}} \geq T_{\mathrm{vir,steady}}$, it may begin steady PopII SF following \cite{Furlanetto22} assuming a PopII SF efficiency of $\eta_{\mathrm{II}} = 0.0139$, which is notably $\sim5.5$ times the F25-Fid PopII SF efficiency. 

Note that for the simulation framework presented here, we apply Eqns. \ref{EQ:Bursty_II} and \ref{EQ:t_quies} deterministically, i.e. every halo is defined by the \textit{mean} relationship and we ignore the log-normal scatter presented in \citetalias{Hazlett25a} and \citetalias{Hazlett25b}. This was done so that our subgrid NN emulators may more accurately reproduce the expected SF for a given SAM framework. However, given its log-normal distribution, we find that the overall effect of including this scatter in our framework is an increase in the high-$z$ PopII SFRDs. To test this, we randomly sampled 100 cells from our fiducial simulation and introduced the scatter in SF between halos. Averaging their SF histories together, we find that PopII SF increases by a factor of $\sim3$ at higher redshifts ($z\gtrsim20$) when the scatter is included, meaning that our fiducial simulation represents a conservative estimate for the PopII SF, and may delay some features of our 21-cm signal predictions which we discuss further in \S \ref{sub:global_results}.\footnote{Note that works such as \cite{Nikolic24} have shown that neglecting stochasticity lengthens the duration of Reionization, especially at high redshifts. We find that if the scatter of \citetalias{Hazlett25b} is included, then our \Tb\ absorption trough is only widened by $\lesssim20\%$.}.

We also incorporate the PopIII SF prescription of \citetalias{Hazlett25b} into our updated fiducial simulation alongside their two-phase PopII SF model. This prescription is also calibrated to \emph{AEOS}, and randomly samples PopIII stellar masses from a cumulative distribution function (CDF, see their Eqn. (5) and accompanying text) which describes the total number of PopIII stars to form within a halo as it overcomes \Mcrit, and is independent of halo mass. Similar to the PopII SF model, we only use the mean relation for PopIII SF, though in this case the mean is estimated via averaging 1000 realizations of the CDF sampling. This results in four PopIII stars per halo, of $M_{\star} = 35.6$ \Msun\ each. To more accurately simulate this updated PopIII SF scenario we adjust properties of the PopIII stars, namely their lifespan and ionizing photon production rates, following the fits presented in \cite{Schaerer02}. This updated PopIII prescription will therefore see longer lived stars and fewer ionizing photons per stellar baryon than the 200 \Msun\ PopIII stars of the ``F25-Fid'' model.

Because the goal of this work is to provide novel 21-cm signal predictions for current and future observational surveys, we compare the results of our fiducial SF prescription to various alternative models to see how differing assumptions alter the 21-cm signal evolution. In what follows, we detail these other alternative models and refer the reader to Table \ref{Table1} for a summary of the various models presented throughout this work. 

Since the exact distribution of PopIII stellar masses (or initial mass function -- IMF) remains largely unconstrained, we assess the impact of PopIII SF on the 21-cm signal by varying the primordial stellar mass introduced within halos as they first overcome \Mcrit. This is achieved by the ``Fid-100'', ``Fid-200'', and ``Fid-400'' simulations, which respectively introduce a single PopIII star of mass 100 \Msun, 200 \Msun, and 400 \Msun\ within halos overcoming \Mcrit. As with the fiducial model, we adjust the properties of the PopIII stars introduced in these simulations following \cite{Schaerer02} so that lower (higher) mass PopIII stars will have longer (shorter) lifetimes and produce fewer (more) ionizing photons per stellar baryon, potentially impacting the 21-cm signal. PopII SF then proceeds within these models as in the fiducial, following \citetalias{Hazlett25a} and \citetalias{Hazlett25b}.

To follow the calibrated SF model of \citetalias{Hazlett25b} more closely, we also include a ``Hi-\Mcrit'' model which uses a boosted critical mass threshold. The authors of \citetalias{Hazlett25b} found that a higher \Mcrit\ with a flatter $z$-evolution than that of \cite{Kulkarni21} resulted in SF histories more closely aligned with hydrodynamic simulations (see Eqns.~(1)-(3) of \citetalias{Hazlett25b} for more detail). Here, we ignore the flattening of \Mcrit($z$) as the authors of \citetalias{Hazlett25b} were focused on regions with no relative baryon-DM streaming velocity (i.e. \vbc\ = 0 km/s), whereas our simulation volume sees velocities as high as \vbc\ $\simeq$ 66 km/s. We find that flattening (i.e. reducing the power law index of) the \cite{Kulkarni21} \Mcrit\ threshold results in a steady \emph{decrease} of \Mcrit($z$) in regions with \vbc\ $\gtrsim$ 45 km/s. Finding this behavior to be unphysical, we choose to only increase \Mcrit\ by a factor of 2.25 to provide a more consistent comparison between the results of our work and \citetalias{Hazlett25b}. 

Similarly, we also include a ``Delayed'' model which lengthens the delay period between PopIII and PopII SF to follow \citetalias{Hazlett25b} more closely. For this, we increase our \tdelay\ parameter from 10 to 100 Myr, which greatly delays the onset of enriched PopII SF within our simulation. The period of time that halo gas requires to recover from PopIII SF and allow for PopII SF is largely unconstrained, but \citetalias{Hazlett25b} found that longer \tdelay\ (up to 300 Myr) yielded SF results that more closely aligned with hydrodynamic simulations. Since our simulation ends at $z = 15$, corresponding to $t_{\mathrm{H}} \simeq 280$ Myr, we adopt \tdelay\ = 100 Myr for the Delayed model as an intermediate assumption, aligning with the findings of other recent works \citep[e.g.][]{Magg18, Magg22, GesseyJones22}. 

As a final comparison, we include the ``HMF Integral'' model of \citetalias{Paper2}, which analytically integrates the HMF to determine the fraction of halo gas which has collapsed and may undergo SF. The change in this collapse fraction over time may then be used to estimate the SFRD of the simulation volume, and this method has been used to rapidly simulate SF in several recent frameworks \citep[e.g.][]{Mesinger11-21cmFAST, Visbal15a, Visbal15b, Jaacks18, Munoz22, Munoz23}. By integrating the HMF over discrete halo mass ranges for PopIII and PopII star-forming halos, this method allows for a rapid analytic estimate of the rise of the first stars, but necessarily lacks the scatter in DM halo merger histories which merger-tree-based simulations include. We therefore implement this as an alternative to the several NN-based subgrid SF models discussed above.

\subsubsection{Neural Network Emulation} \label{subsub:NN_Models}
Our simulation framework was developed (in-part) to mitigate the computational limitations of theoretical models of CD. While \citetalias{Paper1} presented a novel, global SFRD SAM, \citetalias{Paper2} enhanced this framework by allowing for spatial fluctuations within a finite volume, becoming the first large-scale, semi-numeric simulation to include overdensity-dependent DM halo merger histories on small distance scales, self-consistent radiative feedback operating on large scales, and a subgrid SF model which considers \Htwo\ self-shielding and is calibrated to the results of hydrodynamic simulations. This was achieved through the implementation of NN emulators, which decreased the SF computation time by a factor of $\sim$200, yielding self-consistent predictions for the SFRD across CD in $\sim$96 CPU-hours, and now serves as the foundation for the updated simulations of this work.

Since \citetalias{Paper2} mainly served as a proof-of-principle that such NNs may be used to rapidly and accurately emulate the SF of a more complex SAM, we adopt the same fiducial NN architecture for this work (see referenced Table (1)) with minor adjustments to improve its efficiency and thus allow for multiple realizations for comparison. For example, while the PopIII NN framework in this work is identical to that of \citetalias{Paper2}, we have changed the PopII NN output from the stellar mass density to the cumulative stellar mass of PopII stars. This change unifies the output of both PopIII and PopII NNs to be cumulative stellar mass, and provides the PopII NNs with an output value that only increases, making it smoother and naturally easier to emulate.

Beyond this, the models presented here that utilize the two-phase PopII SF prescription of \citetalias{Hazlett25b} necessitated two separate PopII NNs to emulate each phase. The NN architecture for the bursty phase (i.e. for halos with $T_{\mathrm{vir}} < T_{\mathrm{vir,steady}}$) is identical to the fiducial PopII NN emulators described above. The steady phase, however, only applies to merger histories that contain halos more massive than $T_{\mathrm{vir,steady}}$, and we find that including an additional NN input of the logarithm of the most massive halo (MMH) mass, i.e. $\log_{10}(\mathrm{MMH}(z))$, improves the accuracy of these steady-phase PopII SF emulations.

We have also improved the efficiency of our NN training process by reducing the number of SAM realizations required to train each NN. The NNs of \citetalias{Paper2} were trained on the SF results of 1118 combinations of \JLW($z$) and \vbc. Here, we have reduced the number of training \JLW\ histories from 86 to 75, and the number of training \vbc\ bins from 13 to 10, resulting in 750 total combinations which are fed through the SAM to produce the training data for each overdensity bin. 

The 75 training \JLW\ histories were generated from the fiducial model of \citetalias{Paper2} by adding noise to various realistic \JLW\ histories selected to encapsulate the dynamic range of LW backgrounds experienced by the simulation cells. For \vbc, we defined training bins in a manner similar to our overdensity binning procedure, discretizing the distribution of \vbc\ values into bins such that the number of cells within each bin was roughly equivalent. These changes to the \citetalias{Paper2} training procedure reduced the number of training SAM realizations by a $\sim$third while still providing the necessary data to maintain the same emulation accuracy to within 1\%. Overall, while \citetalias{Paper2} utilized $\sim$32,000 CPU-hours to fully generate the training data and train the NNs, our updated framework reduces this to $\sim$24,000 CPU-hours. Lastly, we reduced the runtime of our simulation framework by $\sim$80\% by removing redundancies in the code that caused an unnecessary and repeated loading in of NN data at each redshift step, meaning our fiducial SFRD simulation may be realized in $\sim$21 CPU-hours.

\subsection{The 21-cm Signal} \label{sub:21cm_Model}
The cosmological 21-cm signal results from the hyperfine spin-flip transition of neutral atomic hydrogen in the IGM \citep[e.g. see][for reviews]{FurlanettoAR06, Pritchard12, Liu-Shaw20}, and is typically quantified in terms of its differential brightness temperature with respect to the cosmic microwave background (CMB) at location $\vec{x}$ and redshift $z$, i.e.
\begin{equation} \label{EQ:T_b}
    \delta T_{\rm b}(\vec{x}, z) = \frac{T_{\mathrm{S}}-T_\gamma}{1+z} (1-e^{-\tau}) ,
\end{equation}
where $T_\gamma(z) = (1+z)\times 2.725\ \mathrm{K}$ is the temperature of the CMB, \TS$(\vec{x}, z)$ is the spin temperature of the IGM gas, and $\tau(\vec{x}, z)$ is the optical depth of 21-cm photons through a cloud of hydrogen, given by Eqn.~(15) of \cite{FurlanettoAR06} (see also~\citealt{Dai2019}). \footnote{Note, as with the overdensity and \vbc, we assume all temperatures and $\tau(\vec{x}, z)$ are constant throughout a given cell at $\vec{x}$.} 

\subsubsection{The Spin Temperature} \label{subsub:T_S}
The crux of the 21-cm calculation is the spin temperature, which quantifies the relative number densities of hydrogen atoms within the two hyperfine states. Calculations of \TS, however, are complicated due to its dependencies on the kinetic temperature of the IGM, \TK, as well as the local cosmic overdensity, expansion rate, and Lyman-$\alpha$  (\Lya) background radiation intensity. Throughout this subsection, we follow the form presented in \cite{Hirata06} (henceforth \citetalias{Hirata06}), i.e:
\begin{equation} \label{EQ:T_S}
    T_{\mathrm{S}}(\vec{x},z) = \frac{T_{\gamma}^{-1} + x_{\alpha} T_{\mathrm{c}}^{-1} + x_{\mathrm{c}} T_{\mathrm{K}}^{-1}}{1 + x_{\alpha} + x_{\mathrm{c}}}
\end{equation}
where \TC\ is the effective color temperature (see referenced Eqn. (32) for a detailed description), \xa\ is the Wouthuysen-Field coupling coefficient due to the scattering of UV photons \citep{Wouthuysen52, Field59}, and \xc\ is the coupling coefficient for collisional excitations of hydrogen atoms, given by Eqn. (9) of \citetalias{Hirata06}.

\subsubsection{\texorpdfstring{Lyman-$\alpha$ Background}{subsub:Ly-a}} \label{subsub:Ly-a}
In order to evaluate Eqn. \ref{EQ:T_S} for all simulation cells, we must determine the values of \TK, \TC, and \xa\ for each. The latter two parameters depend on the \Lya\ background intensity seen by the cell, \Ja$(\vec{x},z)$, which we evaluate using Eqn. (55) of \cite{Mittal-Kulkarni21}\footnote{Note that throughout this work, all reported values of \Ja\ have units of [\Ja] = $\rm erg\ s^{-1}\ cm^{-2}\ Hz^{-1}\ sr^{-1}$.} (see their \S 2.3 for further detail).

We assume separate broken power-law SEDs for PopII and PopIII SF, derived to match the assumed number of Lyman-series photons emitted per stellar baryon reported in \cite{Barkana-Loeb05}, i.e. for PopII, $N_{\alpha,\beta}$ = 6520 and $N_{\beta,\infty}$ = 3170 while for PopIII, $N_{\alpha,\beta}$ = 4800 and $N_{\beta,\infty}$ = 2670. The PopII SED is given by Eqn. (49) of \cite{Mittal-Kulkarni21}, and we derive the following SED for PopIII SF:
\begin{equation} \label{EQ:SED_III}
    \epsilon_{\mathrm{b}}(E) = \left\{ \begin{array}{c} 2691.91 (E/E_{\infty})^{0.29} \\ 1155.98 (E/E_{\infty})^{-6.89} \end{array} \mbox{~for~} \begin{array}{c} E_\alpha \leq E \leq E_\beta \\ E_\beta < E \leq E_{\infty} \end{array}\right. 
\end{equation}
where [$E_\alpha$, $E_\beta$, $E_\infty$] = [10.2, 12.09, 13.6] eV are the energies corresponding to the \Lya, Ly-$\beta$, and Ly-$\infty$ transitions of hydrogen.

\subsubsection{The Wouthuysen-Field Effect} \label{subsub:WF-effect}
Once we have determined \Ja\ for all cells, we may then determine the strength of \Lya\ coupling following \citetalias{Hirata06}. With a sufficiently intense \Ja, resonant scattering of Lyman-series photons in the IGM couples \TS\ to the lower \TK\ through  the Wouthuysen-Field (W-F) effect \citep{Wouthuysen52, Field59}. This process lowers \TS\ and therefore \Tb\ (Eqn. \ref{EQ:T_b}) prior to significant X-ray heating, resulting in a \Tb\ absorption trough which is a key prediction for 21-cm cosmology. 

The strength of this effect is quantized by the \xa\ coupling parameter in Eqn. \ref{EQ:T_S}, and is given by Eqn. (11) of \citetalias{Hirata06}. Once we have determined \Ja\ for all cells, we may use it to determine the strength of the W-F effect via this equation, then determine the effective color temperature in by solving
\begin{equation}
    T_{\mathrm{C}}^{-1} = T_{\mathrm{K}}^{-1} + 0.405535\ T_{\mathrm{K}}^{-1} (T_{\mathrm{S}}^{-1} - T_{\mathrm{K}}^{-1}) ,
\end{equation}
which reproduces the the Fokker-Planck equation values to 1\% (\citetalias{Hirata06}). 

It is important to note that the above methodology sees both \xa\ and \TC\ depend on \TS, which itself depends on \xa\ and \TC\ in Eqn. \ref{EQ:T_S}. In order to evaluate \TS\ then, one must iteratively compute \xa\ and \TC\ for some initial guess for \TS\ (typically $T_\gamma$ is a reasonable first guess), and update the \TS\ value with Eqn. \ref{EQ:T_S} until converged. Fortunately, \TS\ converges rapidly in this calculation, typically requiring fewer than five iterations to converge to within 1\%.

\subsubsection{Kinetic Temperature of the IGM} \label{subsub:T_K_IGM}
The final ingredient needed to calculate \TS\ is the kinetic temperature of the IGM gas, \TK, which we evaluate for each cell alongside the free electron fraction, \xe, by solving equations (10) and (11) of  \cite{Mesinger11-21cmFAST}\footnote{Note that we have assumed a neutral fraction of unity in our simulations, and so we do not include the ionization rate term in our calculation of \xe. This is appropriate given our focus on $z>15$, at which times ionized bubbles are much smaller than our cell size \citep[<<1 pMpc,][]{Hayes23}}. We also determine the overdensity of each cell at $z^{\prime}$ by linearly extrapolating their $z=15$ values, an appropriate assumption for our (3 Mpc)$^3$ cells at high redshifts.

To determine the impact of heating, we include the effects of Compton heating due to the scattering of CMB photons off of free electrons and heating due to X-ray emission. We ignore the effects of \Lya\ heating due to scattering as this typically requires large \Ja\ fluxes and so it is most relevant at lower redshifts than those considered here \citep[e.g.][]{Nebrin25}. 

\begin{figure*}
    \centering
    \includegraphics[width=0.8\linewidth]{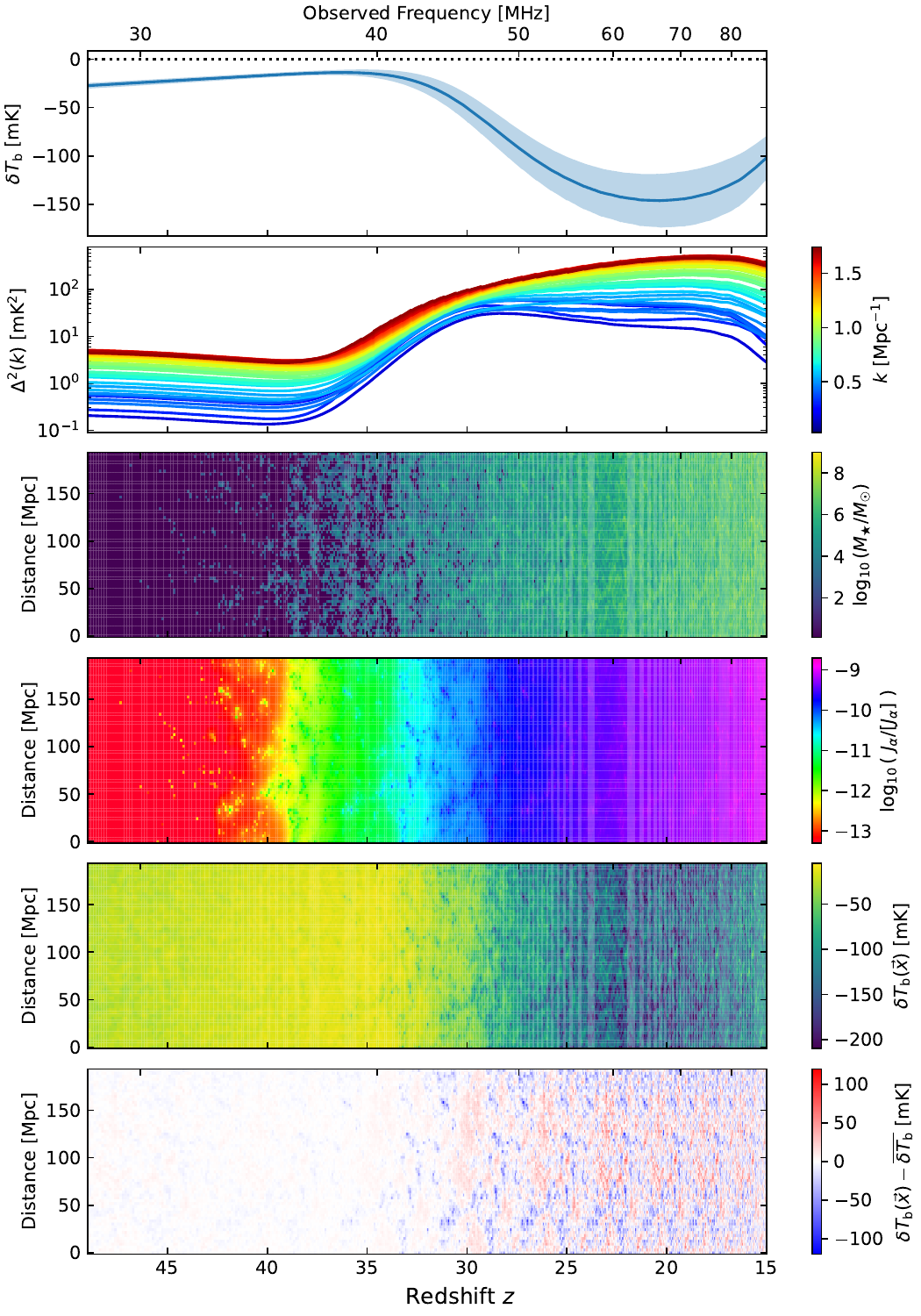}
    \caption{Results of our fiducial simulation framework. \textit{Top panel}: The average 21-cm brightness temperature, $\overline{\delta T_b}(z)$, with 1$\sigma$ standard deviation. \textit{Second panel}: Redshift evolution of the 21-cm power, \PS, for various $k$-modes. \textit{Third panel}: A light cone of the cumulative total (PopII + PopIII) stellar mass formed. \textit{Fourth panel}: A light cone of the Lyman-$\alpha$ background intensity, \Ja. \textit{Fifth panel}: The observed 21-cm light cone. \textit{Bottom panel}: The corresponding differential with respect to $\overline{\delta T_b}(z)$. In all panels, the top axis shows the observed frequency of the signal while the bottom axis denotes the cosmological redshift.}
    \label{fig:Lightcone}
\end{figure*}

We calculate the Compton heating rate per baryon via Eqn. (12) of \cite{Mesinger11-21cmFAST}, and while this process is responsible for coupling \TK\ to $T_\gamma$, its strength grows weaker with time. At lower redshifts, the X-ray emission from intense SF, supernova remnants, and miniquasars becomes the dominant heating mechanism. For our purposes of determining the high-$z$ 21-cm signal, we assume that contributions from remnants and miniquasars trace the star formation rate (SFR). This assumption is appropriate considering the uncertainties in the X-ray emission rate and SED for PopIII stars. The X-ray heating rate per baryon, $\epsilon_{\mathrm{X}}$, is then calculated via Eqn. (17) of \cite{Mesinger11-21cmFAST} (see referenced Section 3.1.2 for further detail). 

We consider 500 evenly spaced bins for X-ray photon energy, ranging from $h\nu_{\mathrm{X}} = 0.1-30$ keV. The SFRD results of each simulation cell are translated into values of $\epsilon_{\mathrm{X}}$ which, alongside the Compton heating rate, are used to obtain \TK\ and \xe. Once we have \TK\ and \xe, we may then combine it with the results of our semi-numeric simulation by calculating \Ja\ and iteratively determining \TS\ (Eqn. \ref{EQ:T_S}) for all cells simultaneously. This is is then used to calculate \Tb\ for all cells via Eqn. \ref{EQ:T_b}.


\section{Fiducial Results} \label{Fid_Results}
Figure \ref{fig:Lightcone} shows the resulting 21-cm signal of our fiducial simulation framework with subgrid SF calibrated to the \emph{AEOS} and \emph{Renaissance} simulations as in \citetalias{Hazlett25b}. The top panel shows the average $\overline{\delta T_b}(z)$ of our simulation with a $1\sigma$ standard deviation. At $z \gtrsim 36$, $\overline{\delta T_b}$ steadily rises with very little scatter as the IGM is largely in the regime of Compton heating, dominated by CMB photons scattering off free electrons, leading to a relatively uniform rise in \TS\ and therefore \Tb. 

The signal reaches a maximum of $\overline{\delta T_b} = -13.79\ {\rm mK}$ at $z = 36.15$ ($\nu_{\mathrm{obs}} \sim\ $ 38.2 MHz) before the \Lya\ feedback from intense SF begins driving \TS\ down to the lower \TK\ via the W-F effect. Because the subgrid cells experience a spread in SF histories (i.e. due to spatially varying \overdense\ and \vbc),
this sources spatial fluctuations in \Ja\ and thus coupling timescales, broadening the standard deviation on $\overline{\delta T_b}$ as it decreases. It reaches an absorption trough minimum of $\overline{\delta T_b} =$ -146.32 mK at $z = 20.25$ ($\nu_{\mathrm{obs}} \sim\ $ 66.8 MHz). After this, X-ray heating begins to drive up \TK, consequently bringing the coupled \TS\ and thus \Tb\ with it. This process begins to wash out spatial fluctuations, narrowing the scatter and resulting in a final prediction of $\overline{\delta T_b}(z=15) = -101.9\ {\rm mK}$ for our fiducial framework.

Such behaviors are mirrored in the second panel of Figure \ref{fig:Lightcone}, which shows the 21-cm power evolution on various $k$-modes. At $z \gtrsim 36$, power steadily decreases as the universe approaches a more uniform temperature. Then, following the onset of SF, power rises on all scales as scattered regions begin to emit \Lya\ radiation and commence W-F coupling. This occurs at different times and rates for different regions of the volume, and so fluctuations rise on all scales. As $\overline{\delta T_b}$ nears its minimum, power begins to plateau on all scales and even decrease at $z \lesssim 17$ due to the rise in X-ray emission/heating which washes out such fluctuations.

The remaining four panels of Figure \ref{fig:Lightcone} respectively show light cones for the cumulative total stellar mass ($M_{\star}$), $J_{\alpha}(\vec{x})$, $\delta T_b(\vec{x})$, and the differential of $\overline{\delta T_b} - \delta T_b(\vec{x})$. These panels further illustrate the findings of the top two panels, as we see the onset and rise of early SF which produces and increasingly intense \Lya\ background. This then overtakes the Compton heating which dominated \TS\ at higher redshifts, and powers the W-F effect which drives \Tb\ down at $z\simeq 35-20$. It is during this period where we see enhanced deviations from $\overline{\delta T_b}$ in the bottom panel. From then on, as the simulation nears $z = 15$, sufficient X-ray emission drives the entire volume up to higher \Tb\ with less intense fluctuations.

\section{Comparison with Alternate Models} \label{Results_Compare}
\subsection{Mean Emissivities} \label{sub:avg_compare}
\begin{figure}
    \centering
    \includegraphics[width=\linewidth]{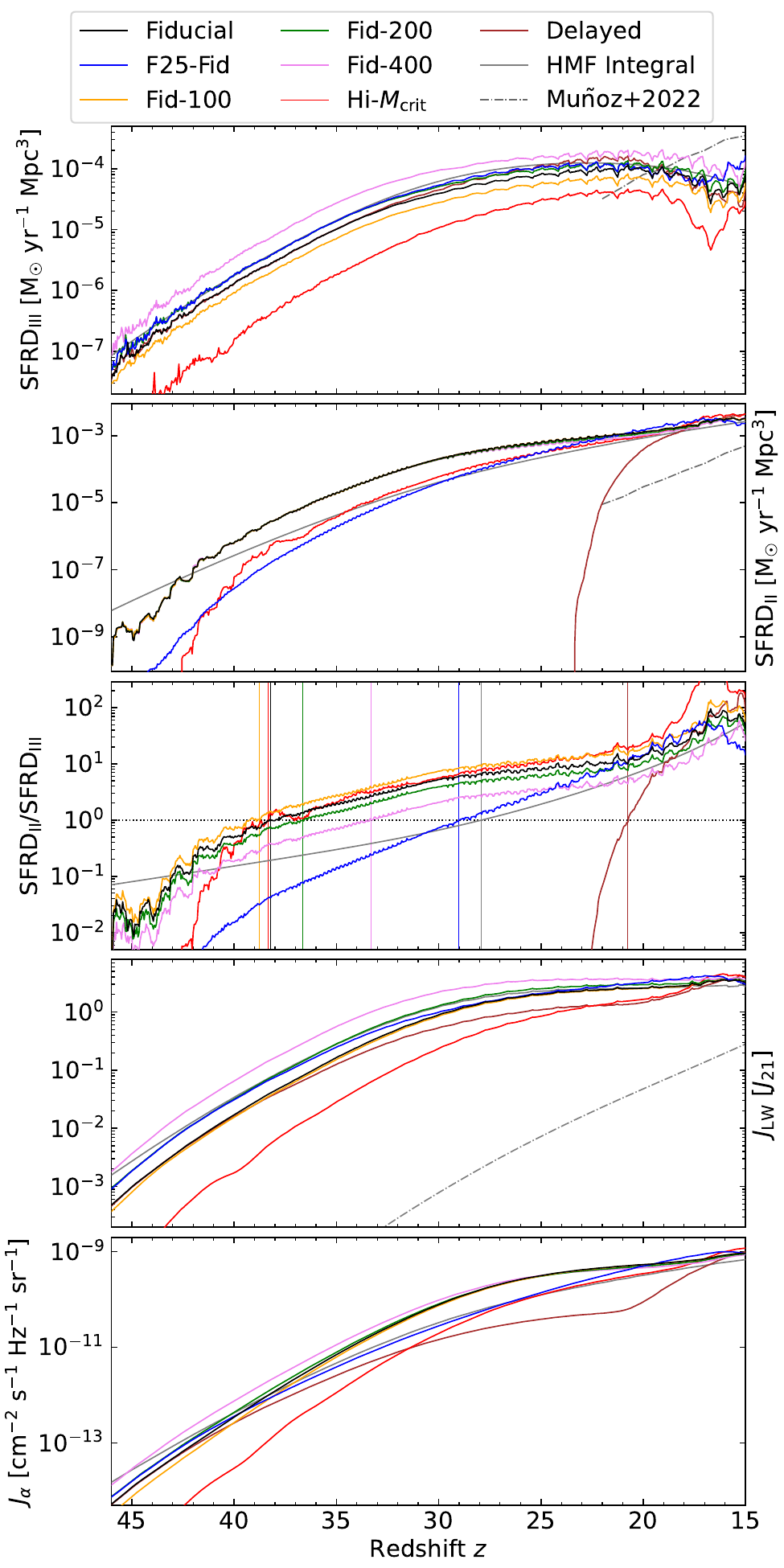}
    \caption{A comparison of the average results of each simulation framework. From top to bottom, panels compare the resulting PopIII SFRD, PopII SFRD, the ratio of the two SFRDs, the LW background intensity (\JLW), and the \Lya\ background intensity (\Ja) averaged across all cells. Within the SFRD ratio plot, vertical dotted lines correspond to the redshift at which each simulation experiences the transition from PopIII- to PopII-dominated SF. For comparison, we also show SFRD and \JLW\ results from the 21cmFAST realization of \protect\cite{Munoz22} (gray dot-dashed).}
    \label{fig:Avg_Results}
\end{figure}

We now compare the results of our fiducial simulation framework to the alternative models described in \S \ref{sub:SF_Model} (Table \ref{Table1}). In Figure \ref{fig:Avg_Results}, we compare the average results of each simulation framework to better understand their underlying differences and to aid in our understanding as to how various simulation assumptions and techniques affect the resulting 21-cm signal. 

The top two panels of Figure \ref{fig:Avg_Results} respectively depict the average PopIII and PopII SFRDs. Focusing on the PopIII, we find that our alternative models of PopIII SF largely scale the resulting SFRD by a constant factor corresponding to the change in PopIII stellar mass introduced within halos. We therefore find that even in large-scale semi-numeric simulations which include DM halo merger histories, the average PopIII SFRD tends to scale linearly with the PopIII stellar mass introduced at each SF event assuming all else is equal. 

The Hi-\Mcrit\ model naturally experiences a delay in the onset of PopIII SF, leading to a systematically lower \SFRDIII\ than any other model. While it is a dex lower than our fiducial SFRD at $z \gtrsim 40$, it steadily rises to only a factor of $\sim$two lower at $z=15$ due to its lower \JLW\ (and thus lower \Mcrit) allowing for more PopIII SF at later times. Conversely, the HMF Integral model was previously calibrated to match the SFRD results of \citetalias{Paper1}, meaning it closely traces the \SFRDIII\ of the F25-Fid and Fid-200 models which each introduce 200 \Msun\ per PopIII SF event, though its evolution is far smoother due to the lack of DM halo merger histories. 

Now looking to the second panel of Figure \ref{fig:Avg_Results}, the Fid-100, Fid-200, and Fid-400 models yield PopII SFRDs that closely match the fiducial to within $\sim10 \%$ across all redshifts shown. This is because the main differences between these four frameworks lies in their PopIII SF models, which are largely disconnected from the PopII. Conversely, the F25-Fid model shows a significant departure from these PopII SFRDs as it does not include a bursty phase prior to steady PopII SF, and utilizes a SF efficiency $\sim5.5 \times$ lower than that of our fiducial model (\citetalias{Hazlett25b}). This causes the fiducial \SFRDII\ to be $\sim$two dex higher than the F25-Fid model at $z \gtrsim 40$, though it grows more slowly afterwards due to the quiescent periods and a lower available halo gas budget such that the two models agree to within $10\%$ at $z \lesssim 22$.

\begin{figure*}
    \centering
    \includegraphics[width=\linewidth]{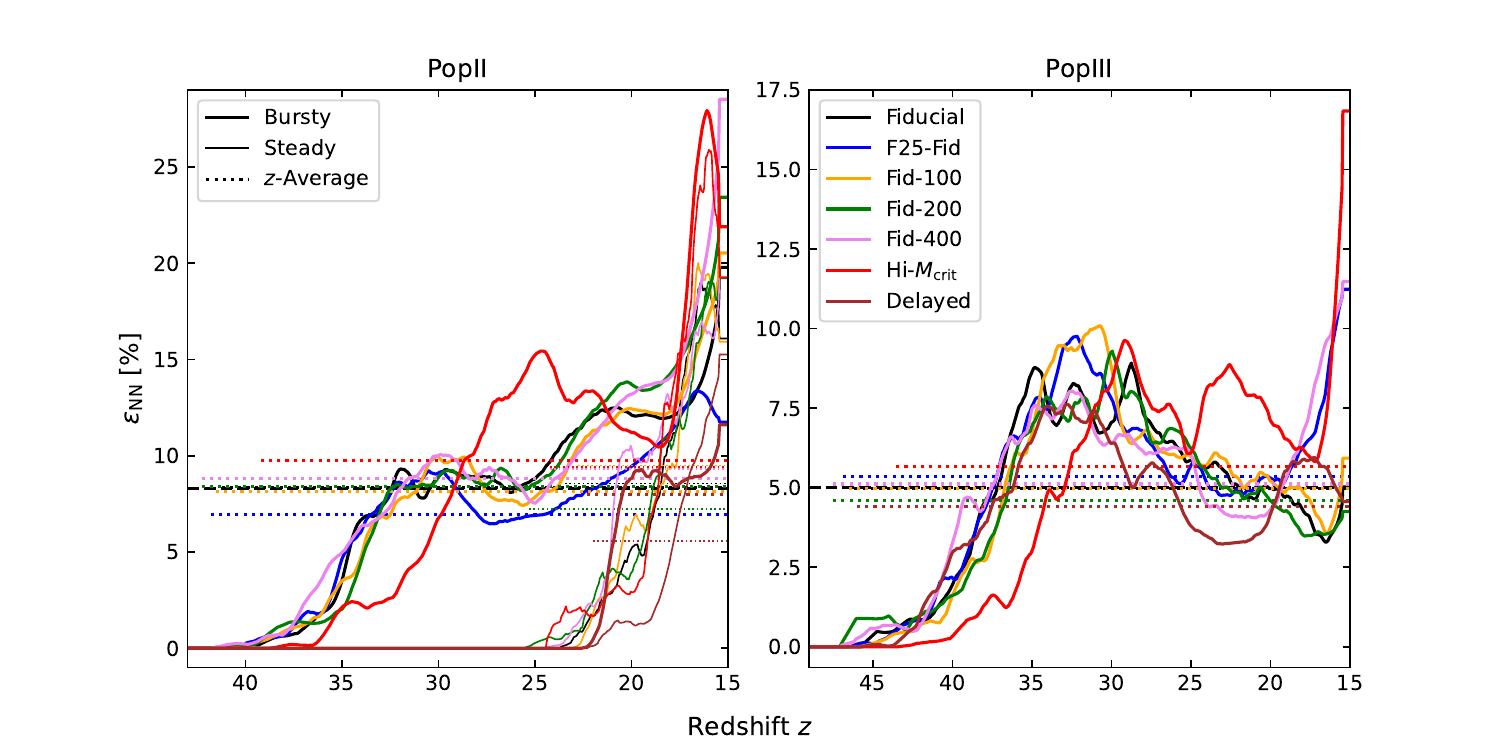}
    \caption{The average error of our NNs by simulation framework as a function of redshift. Each curve represents the average error of 100 randomly sampled cells from the large-scale simulation. \textit{Left panel:} The average error of our bursty (thick lines) and steady (thin lines, if applicable) PopII emulations. \textit{Right panel:} The average error of our PopIII emulations. In both panels, we show the redshift-averaged error of each curve via the corresponding horizontal dotted lines which extend across the redshift range over which each curve is averaged. The horizontal black dashed line in each panel shows the average of the dotted lines, representing the average error of our NN emulators across all redshifts and modelling techniques. For PopII (PopIII), this averaged error is $\sim$8.3\% ($\sim$5.0\%). For clarity, the curves in each panel have been smoothed over an interval ${\mathrm{d}}z = 0.5$.}
    \label{fig:Perc_Error}
\end{figure*}

Now looking to the Hi-\Mcrit\ model, its \SFRDII\ follows a similar trend to its PopIII counterpart in that it is delayed significantly compared to the fiducial due to the higher halo mass threshold for SF. Unlike the PopIII, however, it rises to agree with the fiducial values to within $\sim10\%$ at $z \lesssim 20$. This steeper \SFRDII\ growth agrees more with the F25-Fid model, forming more PopII stars later in the simulation due to higher available halo gas masses than the fiducial at such times. Thus, by adjusting the PopII SF prescription of \citetalias{Paper2} to match that of \citetalias{Hazlett25b}, we see an increase in the high-$z$ \SFRDII. Then, by increasing the \Mcrit\ threshold (as in \citetalias{Hazlett25b}) the high-$z$ \SFRDII\ decreases to yield a similar history to that of \citetalias{Paper2}. 

A similar albeit more extreme trend is then seen in the Delayed model; with a delay period ten times the fiducial value, PopII SF does not begin in this model until $z \sim 23$, after which it rapidly rises to meet the other models to within $\sim10\%$ by $z \sim 18$. Once again, we find that the available halo gas mass heavily constrains PopII SF such that each model naturally yields similar \SFRDII\ values at low redshifts ($z \lesssim 20$), regardless of their assumptions/techniques. 

The third panel of Figure \ref{fig:Avg_Results} shows the ratio of \SFRDII / \SFRDIII, with correspondingly colored vertical lines at the redshift where PopII SF overtakes PopIII. This transition occurs at $z \sim 38$ in our fiducial model, which is similar to the timing for the Fid-100, Fid-200, and Hi-\Mcrit\ models. The F25-Fid and HMF Integral models experience later transitions due to their lower PopII SF efficiencies, while the Delayed model does not see PopII dominate until $z \sim 21$, simply due to its increased delay period following PopIII SF. 

The average \JLW\ of each model in the fourth panel of Figure \ref{fig:Avg_Results} bears a striking resemblance to the PopIII SFRDs, which is somewhat expected as the \JLW\ of each model is dominated by PopIII stellar emission until very late into the simulation due to its higher LW photon production than PopII. It is only at $z \lesssim 20$ where relatively weaker PopII feedback meaningfully contributes, leading to close \JLW\ agreement between models at this time, reflecting the behavior of the PopII SFRDs at this time.

The bottom panel of Figure \ref{fig:Avg_Results} shows the \Lya\ radiative background intensities, \Ja\footnote{Note that \Ja\ of each model evolves in a remarkably similar fashion to the X-ray background intensities, \JX. Our findings for \Ja\ therefore apply to \JX\ as well, though the values of \JX\ are significantly smaller by about eight dex.}. At high redshifts ($z \gtrsim 35$), \Ja\ is largely reflective of the PopIII SFRDs, dominated by their stellar emission. Then, similar to \JLW, it becomes dominated by the PopII SFRD at the lower redshifts such that \Ja\ agrees across all models to within $\sim20\%$ at $z \lesssim 17$. Unlike \JLW, however, this transition to PopII dominating the \Ja\ occurs at roughly the same time as PopII dominating the SFRD (third panel) due to its higher UV SED (Eqns. \ref{EQ:SED_II} and \ref{EQ:SED_III}). Differences in SFRDs (particularly PopII) between models therefore source differences in \Ja\ at these intermediate redshifts which, as we shall see below, manifests as differences in W-F coupling strength and thus \Tb. 

Finally, in Figure \ref{fig:Avg_Results} we have included results from the \textsc{AllGalaxies} model of \cite{Munoz22}, computed via 21cmFAST \citep{Mesinger11-21cmFAST, Murray20} for comparison\footnote{Note that this realization of 21cmFAST was computed with an emphasis on CD and the EoR ($5 < z<35$), and therefore does not provide comparison at $z \gtrsim 35$.}. This model employs separate integrals over the HMF for PopII and PopIII stellar populations, similar to our ``HMF Integral'' method. We find that the \SFRDIII\ of this model is comparable to ours at $z\sim 18-22$, but is much steeper, declining rapidly towards higher redshifts. Conversely, the \SFRDII\ is significantly weaker than our predictions for $z>15$, instead coming to dominate at much later times. 
The lower total SFRD than in our simulations paired with the lower assumed stellar emissivities in \cite{Munoz22} (henceforth \citetalias{Munoz22}) results in far weaker radiative feedback, evidenced by \JLW\ being 2-3 orders of magnitude lower than our model predictions for $z>15$.

\subsection{Emulation Accuracy} \label{sub:NN_accuracy}
We demonstrate the accuracy of each NN framework against the full SAM results in Figure \ref{fig:Perc_Error} (see Fig. 3 of \citetalias{Paper2} and accompanying text for further detail). Across all redshifts, we find that the average NN emulation is accurate to within $\sim 30\%$ and $\sim 17\%$ for PopII and PopIII stars, respectively. In general, the error of our NN emulations increase with decreasing redshift as they largely depend on previously emulated SF, meaning that error tends to accumulate with time, though exceptions exist as several of our PopIII emulators see a decrease in error from $z\simeq 32-17$. 

While this SF error is certainly non-negligible, we find its impact to be subdominant to uncertainties in SF efficiency and feedback at such early times. Error from NN emulation may subtly shift the final SFRD values up or down by tens of percent, though we find that it has minimal impact on the resulting power spectra. Unfortunately, subtle differences in the simulation method and the NN training process make determining the exact source of such error for each individual model challenging. 

On average, a given model will respectively be accurate to within $\sim 10\%$ and $\sim 6\%$ at any given redshift, as denoted by the horizontal dotted lines in Figure \ref{fig:Perc_Error}. Averaging these redshift-averaged values across all models then gives the horizontal dashed black lines in each panel, representing the average error one expects from our NN emulators regardless of redshift or modelling technique. From these, we find that our PopII and PopIII NNs respectively average about 8.3\% and 5.0\% error across all models and redshifts. 

\subsection{Global 21-cm Signal} \label{sub:global_results}
\begin{figure*}
    \centering
    \includegraphics[width=\linewidth]{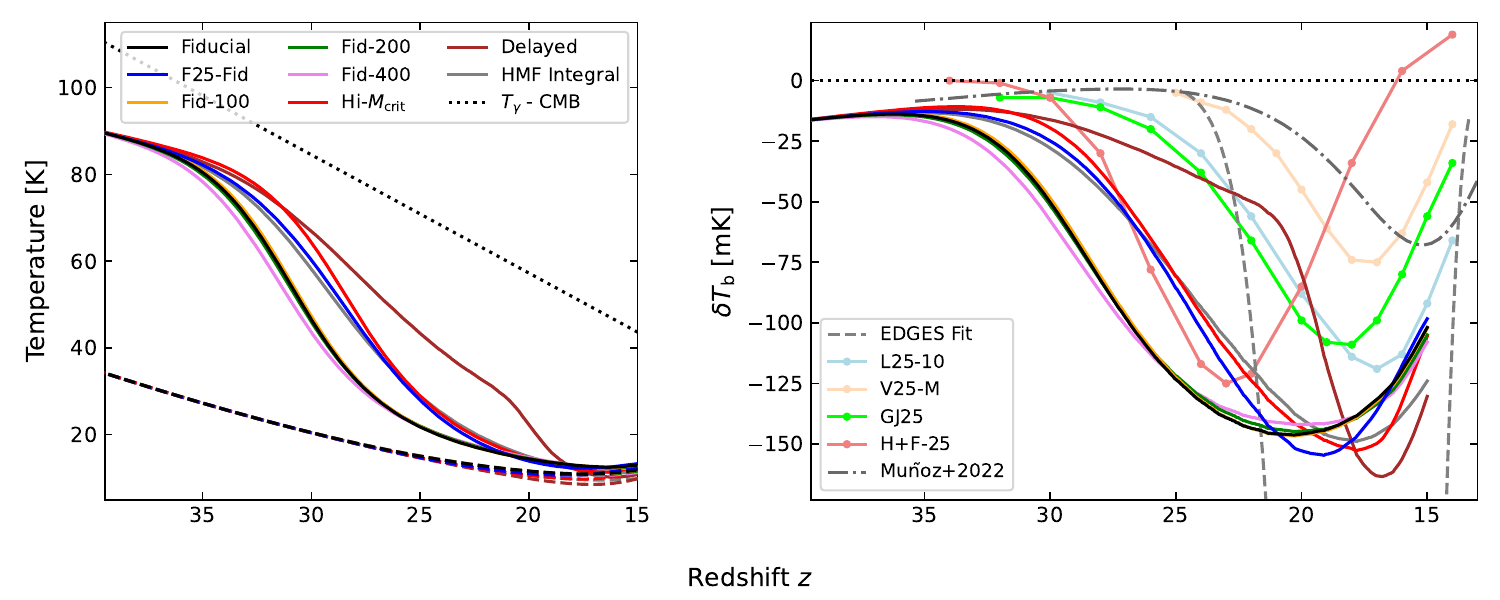}
    \caption{\textit{Left:} A comparison of the average IGM \TS\ (solid) and \TK\ (dashed) resulting from each of our simulation frameworks, with the CMB temperature represented by the dotted black line. \textit{Right:} The corresponding \Tb($z$) compared with previously published 21-cm results. For comparison, we include the claimed EDGES detection \protect\citep[gray dashed,][]{Bowman18-EDGES} as well as various theoretical predictions from \protect\cite{Liu25} for a delay period of 10 Myr (L25-10), \protect\cite{Ventura25} for moderate PopIII SF (V25-M), \protect\cite{GesseyJones25} (GJ25), \protect\cite{Hegde25} (H+F-25), and the \textsc{AllGalaxies} model of \protect\cite{Munoz22}, computed via 21cmFAST (gray dot-dashed).}
    \label{fig:All_21cm}
\end{figure*}

We now turn our attention to the main predictions of our simulation, the 21-cm brightness temperature. In Figure \ref{fig:All_21cm}, we compare the  $\bar{T}_S(z)$ and $\bar{T}_K(z)$ for each of our simulations (left) and their corresponding 21-cm global signal (right). Based on their global spin temperature histories (left panel), we divide our simulations into three groups: (i) Fiducial, Fid-100, Fid-200, and Fid-400, (ii) HMF Integral, F25-Fid, and Hi-\Mcrit\, and (iii) Delayed. We will respectively refer to these as the ``Early-'', ``Intermediate-'', and ``Delayed-'' coupling models throughout the remainder of this work. 

These three groups begin undergoing \Lya\ coupling at different times, and evolve at different rates. The Early-coupling models all peak in \Tb\ at $z > 36$, while the Intermediate- and Delayed-coupling group peak at $z \simeq 34-36$. The rate at which the IGM \TS\ drops after this point is heavily dependent on the evolution of \Ja\ which, as shown in Figure \ref{fig:Avg_Results}, differs most between models at $z = 35-17$. If, for example, we assume \TS\ to be coupled to \TK\ when the two values are within 50\%, the Early-, Intermediate-, and Delayed- coupling groups respectively couple at $z \sim$ 26, 22, and 18. 

The onset of PopII SF causes a sudden decrease in the \TS\ and \Tb\ for the Delayed model at $z \sim 21$. The sharpness  of this transition is somewhat unphysical, however, deriving from the simplistic constant delay period of 100 Myr following PopIII SF for all halos. A more realistic distirbution of recovery periods varying with local conditions would soften this feature. As \tdelay\ is still largely unconstrained, we defer a more accurate representation to future work. Overall, the behavior of \TS\ in the left panel of Figure \ref{fig:All_21cm} mirrors the relative differences in \Ja\ seen in Figure \ref{fig:Avg_Results}.  

At the lowest redshifts ($z \lesssim 18$), \TK\ stops decreasing from cosmic expansion, and begins rising due to X-ray emission in all models. Table \ref{Table2} lists the minimum 21-cm brightness temperature achieved by each model, \Tbmin, and the corresponding redshift at which it reaches this minimum, $z_{\mathrm{min}}$. Interestingly, we find that our Early-coupling models all reach a minimum of $\delta T_{\mathrm{b,min}} \simeq\ -145\,{\rm mK}$ at $z = 20.25$ ($\nu_{\mathrm{obs}} \simeq 66.8$ MHz), suggesting that PopIII SF has minimal impact on absorption depth or timing, though models with lower (higher) PopIII stellar mass reach a slightly deeper (shallower) \Tbmin\ due to their slightly higher (lower) \SFRDII\ at such times. The Intermediate-coupling models are more spread out, yielding $-149\ {\rm mK} \gtrsim$ \Tbmin\ $\gtrsim -155\ {\rm mK}$ at $z_{\mathrm{min}} \sim 18-19$. Here, differences once again manifest due to differences in \SFRDII\ at this time. Finally, the rapid growth of SFRD at $z\lesssim20$ causes the Delayed model to reach the latest and strongest 21-cm absorption of all models. As its SFRD (and consequently, \Ja) reach parity with other models around $z\sim17$, its turn-around due to X-ray heating is sharp.

Putting it all together, we find that an earlier rise in SFRD manifests as an earlier onset of W-F coupling and a shallower \Tbmin\ as increased X-ray feedback results in earlier and more gradual IGM heating. This is most clearly seen by comparing the fiducial \Tb\ with that of the Delayed model -- while their SFRDs at $z = 15$ agree quite closely, the Delayed model experiences far more intense SF at much later times than the fiducial, resulting in relatively weak \Lya\ coupling prior to the onset of PopII SF. Then, as \SFRDII\ rapidly rises, we see accelerated \Lya\ coupling followed by a quick turnaround to X-ray heating, resulting in the latest, narrowest, and deepest absorption trough of all our models. Overall, we find that the relative differences in \Ja, which are heavily dependent on their PopII SFRDs, dominate the behavior of \Tb.

We also compare our results to the claimed EDGES detection \citep{Bowman18-EDGES} in Figure \ref{fig:All_21cm}. While our models are much more shallow ($\sim$1/3 the EDGES \Tbmin), the $z_{\mathrm{min}}$ closely aligns with those of our Intermediate- and Delayed-coupling models. The Early-coupling models predict slightly earlier minima, but all of our $T_{\mathrm{b,min}}$ fall within the redshift range of the claimed EDGES trough detection\footnote{As shown in other works \citep[e.g.][]{Fialkov19, Reis20b, Sikder24}, the extreme depth of the EDGES result may be achieved through a strong radio background, though we defer the inclusion of such a background to future work \citep[also see][]{Ahn21, Cang25}.}.

Figure \ref{fig:All_21cm} also illustrates that our \Tbmin\ are generically early and deep compared to the bulk of previously published models in the literature. The greater depth and earlier timing of our \Tb\ absorption trough largely stems from our enhanced SFRD over these models; however, the physical reasons for this enhancement differs between works. For example, the total SFRD of \cite{Hegde25} is less than half our fiducial SFRD, but the authors ignore the stochasticity of halo merger trees and assumed a uniform IGM temperature, leading to more uniform and continuous \Ja\ coupling and X-ray heating throughout CD, resulting in earlier transition to emission than any other model despite their lower SFRD. 

The other previously published \Tb($z$) predictions in Figure \ref{fig:All_21cm} result from even \textit{lower} total SFRDs, which leads to weaker \Lya\ coupling and shallower \Tb\ minima. The \cite{Ventura25} model does not include minihalos, thereby delaying PopIII and subsequent PopII SF. Similarly, the \cite{Liu25} model assumes a systematically higher \Mcrit\ threshold than that of our model, which again delays PopIII (and then PopII) SF. The \cite{GesseyJones25} model in Figure \ref{fig:All_21cm} assumes a longer \tdelay\ of 30 Myr, bisecting our fiducial (10 Myr) and Delayed models (100 Myr), delaying PopII SF and feedback as well.

\begin{table}
    \centering
    \begin{tabular}{|c|c|c|c|}
        \hline
         Model & Coupling & \Tbmin [mK] & $z_{\mathrm{min}}$ \\
        \hline\hline
         Fiducial & E & -146.32 & 20.25 \\
         \hline
         F25-Fid & I & -154.67 & 19.10 \\
         \hline
         Fid-100 & E & -146.73 & 20.25 \\
         \hline
         Fid-200 & E & -144.94 & 20.25 \\
         \hline
         Fid-400 & E & -142.08 & 20.25 \\
         \hline
         Hi-\Mcrit\ & I & -152.79 & 17.80 \\ 
         \hline
         Delayed & D & -163.42 & 16.80 \\ 
         \hline
         HMF Integral & I & -149.03 & 17.80 \\
         \hline
    \end{tabular}
    \caption{Absorption trough minima for our various semi-numeric models. From left to right, columns detail the model name, the coupling group to which it belongs (E = Early, I = Intermediate, and D = Delayed), the minimum 21-cm brightness temperature, \Tbmin, and the corresponding redshift of \Tbmin. Values obtained from Figure \ref{fig:All_21cm}.} \label{Table2}
\end{table}

With the exception of \cite{Hegde25}, the \Tbmin\ of these comparison models most closely align with our Delayed model. However, we found the knee in the Delayed \Tb($z\simeq20$) to be somewhat unphysical as one expects a distribution of \tdelay\ periods rather than a constant value. Thus, if a more realistic PopIII-II transition model is implemented for our Delayed model, it would allow for an earlier and more gradual transition to PopII domination and \Lya\ coupling, leading to a shallower and wider \Tb\ absorption trough that more closely resembles those shown from the literature as PopII X-rays begin to contribute earlier. 

\begin{figure*}
    \centering
    \includegraphics[width=\linewidth]{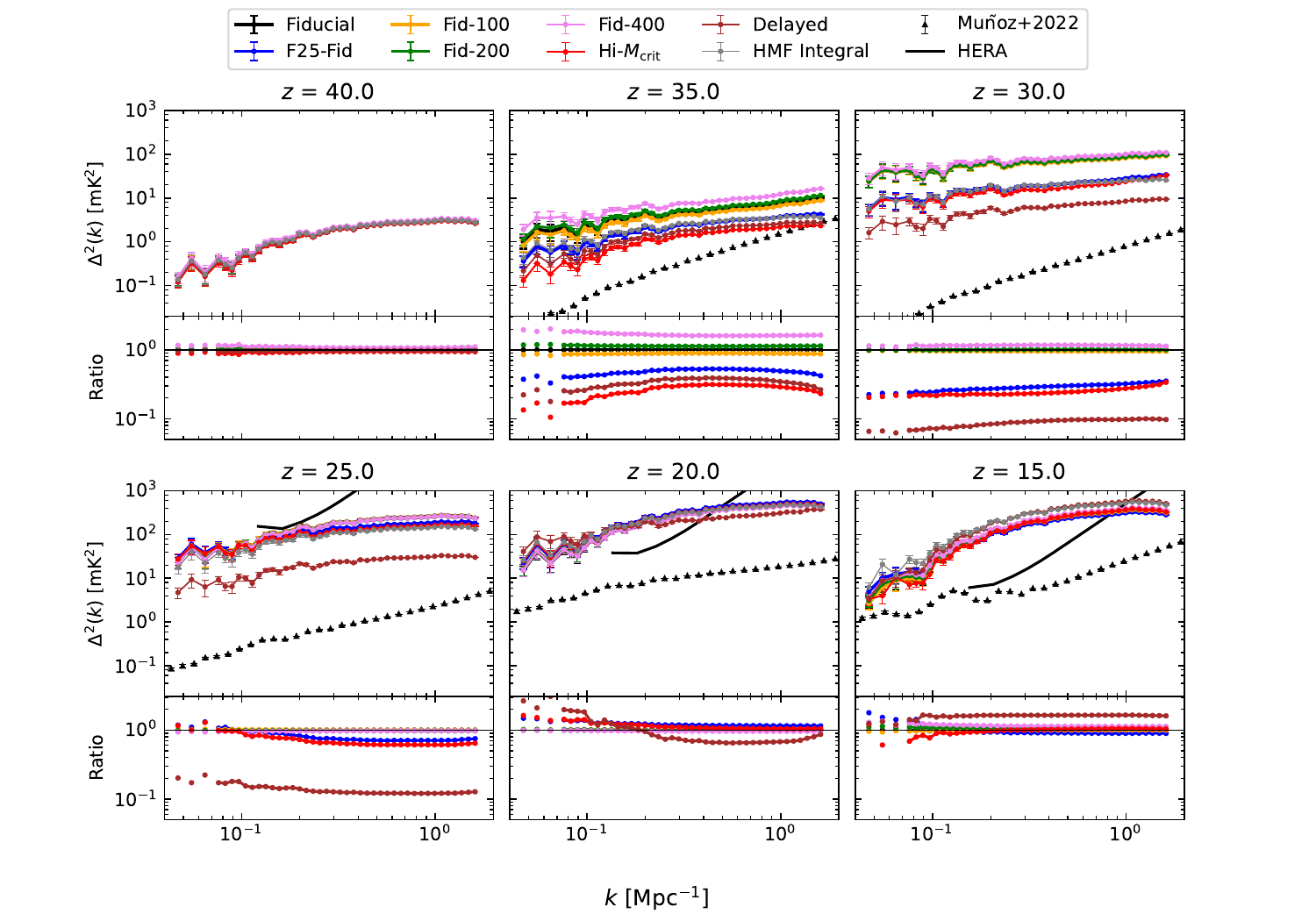}
    \caption{A comparison of the dimensionless 21-cm power spectra resulting from each simulation framework. From top left to bottom right, we compare the resulting \PS\ at various redshifts from $z = 40 - 15$. At each redshift, the top panel shows the power spectra of each semi-numeric simulation, while the bottom panels depict the corresponding ratio of each model with respect to the fiducial. We also show the results of the \textsc{AllGalaxies} model of \citetalias{Munoz22}, computed via 21cmFAST in all $z\leq35$ panels (black triangles). Finally, we also show sensitivity curves for HERA in the bottom row ($z=25-15$) via solid black (see text).}
    \label{fig:Power_Spectra}
\end{figure*}

Finally, the \citetalias{Munoz22} realization of 21cmFAST shows the latest and shallowest \Tb\ absorption trough of all models in Figure \ref{fig:All_21cm}. Given its far lower \SFRDII\ (Figure \ref{fig:Avg_Results}) and assumed stellar emissivities, this realization yields far less intense radiative feedback, resulting in weaker \Lya\ coupling and X-ray heating, translating to \Tbmin $\simeq -67.9$ mK at $z \simeq 15.1$.

In sum, Figure \ref{fig:All_21cm} shows that: \textbf{1.)} The inclusion of minihalos and stochastic DM halo merger histories allows for an earlier onset of SF and subsequent feedback/coupling. \footnote{Further, should the scatter in SF of \citetalias{Hazlett25a} and \citetalias{Hazlett25b} be included (as discussed in \S \ref{subsub:SF_Model}), we find that the resulting increase in \SFRDII\ shifts our 21-cm predictions to earlier times by $\Delta z \sim 3$, and so features of our 21-cm results should be considered conservative lower limits in redshift space.} \textbf{2.)} The consideration for \Htwo\ self-shielding in our \Mcrit\ model similarly allows for earlier SF and feedback. \textbf{3.)} The delay period separating PopIII and PopII SF has a large impact on the resulting 21-cm signal, and since this is only applicable to simulations with DM halo merger histories, further emphasizes the importance of considering realistic DM halo evolutions. 

\subsection{21-cm Power Spectra} \label{sub:PS_results}
\subsubsection{Evolution with scale and redshift}
\begin{figure*}
    \centering
    \includegraphics[width=\linewidth]{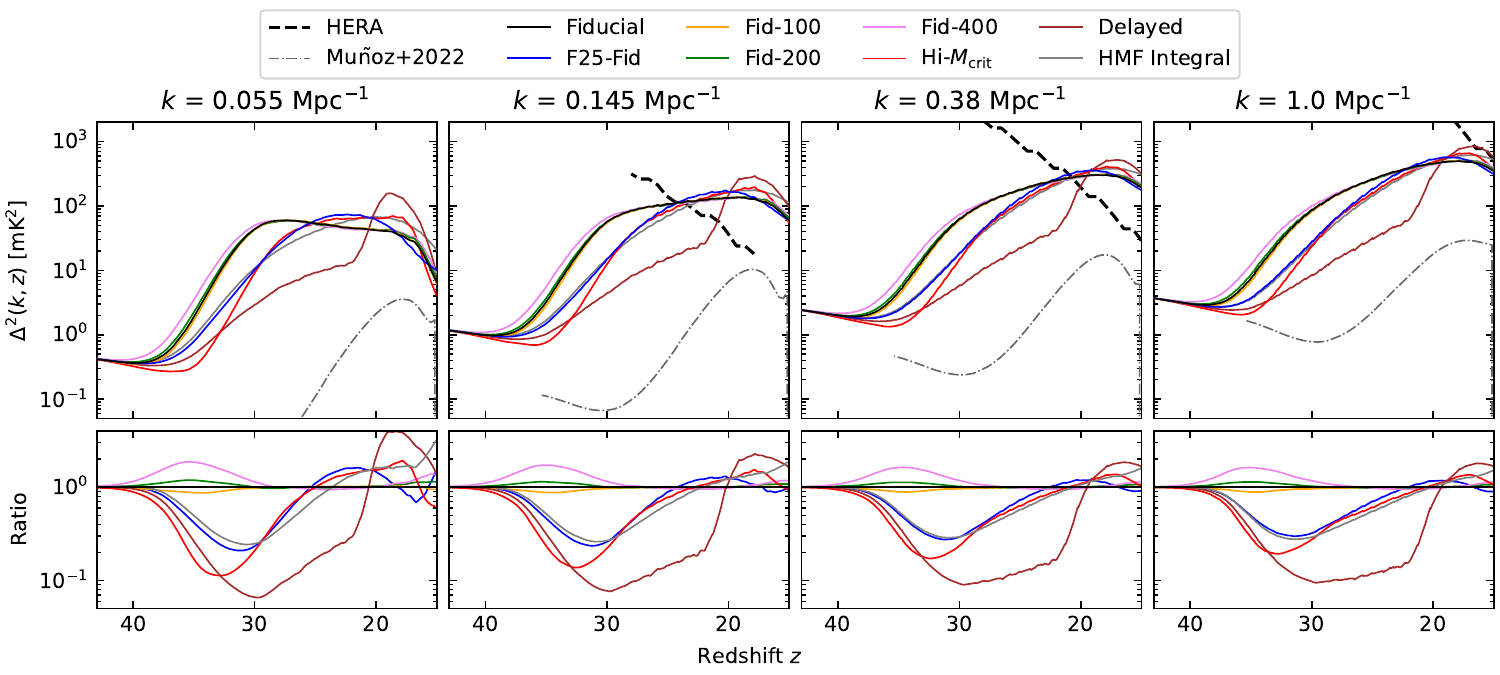}
    \caption{Redshift evolution of the 21-cm \Tb\ power on four various $k$-modes, as denoted by the title. The top subplot of each panel displays the power evolution of each simulation, and the bottom subplot shows the ratio of power with respect to the fiducial model. We also show the power evolutions of \citetalias{Munoz22} \textsc{AllGalaxies} model resulting from 21cmFAST (gray dot-dashed), as well as HERA sensitivity curves obtained from 21cmSense for the three largest $k$-modes (black dashed).}
    \label{fig:k-modes}
\end{figure*}

We present the dimensionless 21-cm power spectra, \PS, of each model as a function of scale $k$ at several redshifts in Figure \ref{fig:Power_Spectra}, and inversely as a function of redshift for several discrete scales in Figure \ref{fig:k-modes}. 
In both, the top panels depict the power, while the bottom panels show the ratio of each model to the fiducial. For comparison, we also show in each the model of \citetalias{Munoz22} for $z\leq 35$.

To first order, all of our models have the same shape as a function of $k$ for any particular redshift, though this shape evolves over redshift\footnote{We again note that, should the SF scatter of \citetalias{Hazlett25a} and \citetalias{Hazlett25b} be included in our framework, evolutionary features of our power spectra would occur sooner by $\Delta z \sim 3$ as a result of increased PopII SF.}. Differences between the models are almost entirely in their amplitudes, and are sourced primarily by differences in SFRD at early times, and by X-ray heating at later times.

At $z \gtrsim 40$, all models are identical as not enough SF has occurred yet to meaningfully affect the IGM. The onset of \Lya\ coupling, however, is clearly seen at $z \sim 35-40$ as a sudden rise in power across all models and all distance scales, with larger scales rising more rapidly due to PopIII domination in regions with relatively low \vbc. By $z = 30$, \Lya\ coupling has grouped the power spectra into the Early-, Intermediate-, and Delayed-coupling groups defined previously, mirroring the behavior seen in Figures \ref{fig:Avg_Results} and \ref{fig:All_21cm}. This indicates that the total SFRD dominates the IGM conditions at intermediate redshifts prior to X-ray heating. 

Between $z = 25-20$ the Intermediate- and Delayed-coupling models rise to the amplitude of the Early-coupling models as their SFRDs rapidly accelerate. By $z = 20$, the power spectra of all Early- and Intermediate-coupling models agree to within 10\% across all scales. The Delayed-coupling model is an interesting exception here: while its power spectrum amplitude remains in broad agreement with the other models, it is modestly but significantly flatter with respect to $k$ (c.f. Figure \ref{fig:Power_Spectra}). That is, it has enhanced large-scale and suppressed small-scale power due to the rapid growth of its SFRD, which enhances fluctuations correlated with \vbc. We will discuss this in more detail shortly. 

By $z = 15$, X-ray heating driven by PopII SF takes hold in all models and it begins to wash out fluctuations on all scales, though it does so more rapidly on larger scales \footnote{Recall that we have assumed $x_{\mathrm{HI}}$ = 1 throughout our simulation. Though ionized bubbles surrounding the first sources may enhance small-scale 21-cm power, the scale of such bubbles are $<<1$ pMpc at $z\gtrsim11$ \citep{Hayes23}. We therefore do not include the effects of reionization as the largest bubbles would exist on scales comparable to our cell size at $z \sim 15$, with radii falling at higher redshifts.}. All models agree to within a factor of $\sim$two across all scales, mirroring the behavior of the PopII SFRDs and \Ja. Here, we find that \textit{relative} fluctuations are similar for all models, and that the amplitude of the power (especially at smaller scales) is driven by the amplitude of the global mean (c.f. Fig. \ref{fig:All_21cm}).

\subsubsection{Comparison with \citetalias{Munoz22}}
Comparing our models to \citetalias{Munoz22}, we notice two main patterns: (i) \citetalias{Munoz22} has generically lower power (seen in both figures), and (ii) their power has a steeper slope at high redshifts, flattening to match our slope by $z\sim20$ (seen in Figure \ref{fig:Power_Spectra}).

To explain (i) we note that \citetalias{Munoz22} adopts a SF prescription similar to our HMF Integral model, meaning there are no discrete DM halos, no delay period between PopIII and PopII SF, and no quiescent periods in between bursts of PopII SF as in our models. These differences combine to yield a more homogeneous 21-cm signal, as every subgrid cell quietly and consistently forms stars throughout CD, which irradiates the volume more uniformly than the bursty SF prescriptions of this work, resulting in lower power spectra. 

To explain (ii), we note that the 21-cm power spectra of \citetalias{Munoz22} at $z\gtrsim30$ closely follows the shape of the underlying matter power spectrum, while our models have significantly enhanced large-scale power due to the effects of \vbc\ on SF; regions of low \vbc\ reduce \Mcrit, resulting in a scale-dependent SFRD that depends on fluctuations in \vbc. These fluctuations are strong on large spatial scales, and therefore strengthen large-scale SFRD fluctuations, and ultimately 21-cm power. This large-scale 21-cm power enhancement also occurs in \citetalias{Munoz22} (note the shape at $z\sim20$), but is delayed  with respect to our models due to their systematically higher \Mcrit\ threshold and relatively simplified SF prescription.

\subsubsection{Detectability with HERA}
We indicate the nominal sensitivity of HERA observations as solid black lines in Figure \ref{fig:Power_Spectra} and dashed black in Figure \ref{fig:k-modes}. Computed via \textsc{21cmSense v2} \citep{Murray24}, we assume 1080 hr of drift-scan observations (6 hours per night over 180 days, averaged coherently within LST chunks of the size of the FWHM of the beam, i.e. $\approx 1\, {\rm hr} \times (150\,{\rm MHz}/\nu) $) with the full split-core HERA-320 array. We assume an achromatic receiver temperature of $T_{\rm rcv}=100$\,K and a sky temperature of $T_{\rm sky}=(\nu/150\ {\rm MHz})^{-2.6}$. We adopt a `moderate' foreground-removal scenario that assumes foregrounds are removed out to the horizon delay plus a buffer of 0.1 $h$/Mpc, and in which redundant baselines are averaged coherently  \citep[see][for details]{Pober14}. The sensitivity at each redshift assumes an 8\,MHz spectral bandwidth split across 81 channels.

Between $z = 25-20$, the \PS\ of all models becomes detectable with HERA, beginning with $k\sim0.2\ \mathrm{Mpc^{-1}}$ and expanding to smaller scales such that by $z=15$ we might detect up to $k\lesssim 0.9\ \mathrm{Mpc^{-1}}$. 

While a detection of the power predicted by our models appears relatively comfortable, the sensitivity required to \textit{distinguish} between our models is much higher. Nevertheless, distinguishing between our model and the \textsc{AllGalaxies} model of \citetalias{Munoz22} will be more straightforward; under these sensitivity assumptions, a detection at $z\gtrsim 15$ will constitute substantial evidence in favor of our model, including discrete DM halo merger histories and calibration to the SF results of hydrodynamic simulations.

Furthermore, the second panel of Figure \ref{fig:k-modes} reveals that this forecast sensitivity may be sufficient to distinguish our ``Delayed'' PopII SF model from the others. This finding comes with the caveat, however, that the sudden rise in PopII SF and power in the Delayed model may be somewhat unphysical in that a more realistic \tdelay\ model may smooth its growth, decreasing our ability to distinguish it from other models. 

\subsubsection{Summary of power spectra}
In sum, Figures \ref{fig:Power_Spectra} and \ref{fig:k-modes} reveal that the effects of PopIII SF dominate power on all scales at high redshifts, while the effects of PopII SF begin to dominate at later times. In the transition period, PopIII tends to dominate larger-scale power, and PopII the smaller-scale power. Furthermore, we find that power on all scales for $z>15$ is enhanced in our model compared to that of \citetalias{Munoz22}, primarily due to our enhanced SF prescriptions. 

\section{Conclusions \& Discussion} \label{Conclusions}
In this work, we built upon the previous semi-numeric simulation of \citetalias{Paper2} to include SF prescriptions that are calibrated to state-of-the-art hydrodynamic cosmological simulations. We achieved this by adopting the SF models presented in \citetalias{Hazlett25a, Hazlett25b}, which calibrate both PopIII and PopII SF to the results of \emph{Renaissance} and \emph{AEOS}. We then extended the simulation of \citetalias{Paper2} to include a post-processing 21-cm \Tb\ calculation, which translates the resulting spatial distribution of stars into predictions for the 21-cm global signal, power spectrum, and their evolution across CD. 

The simulation framework of \citetalias{Paper2} depends on artificial NNs which emulate the SF history of a more complex, self-consistent SAM including stochastic DM halo merger trees (\citetalias{Paper1}). By further improving the efficiency of this simulation framework, SFRD results may be obtained in $\sim$21 CPU-hours once NN training is completed, allowing for multiple realizations to be performed with varying assumptions and modelling techniques to see the impact that each has on the 21-cm signal. 

Overall, we find our NN emulators to be quite accurate, reproducing the SF history of our DM halo merger tree-based SAM with an average accuracy of $\sim8.3\%$ and $\sim5.0\%$ for PopII and PopIII SF, respectively, regardless of redshift or modelling technique (Figure \ref{fig:Perc_Error}). While the accuracy of our models exhibits some redshift dependence (in that it typically grows with time due to the accumulation of emulation errors), differences in model assumptions and subtleties of NN training make it difficult to determine the source of this behavior.

The simulation framework presented in here is the first such large-scale, semi-numeric simulation to include stochastic DM halo merger histories, self-consistent radiative feedback, and prescriptions for both \Mcrit\ and the SF within halos calibrated to hydrodynamic simulations. When comparing our results with previously published 21-cm predictions (Figure \ref{fig:All_21cm}), we find that the combination of DM halo merger trees, the \Mcrit\ model of \cite{Kulkarni21} which accounts for the effects of \Htwo\ self-shielding, and a relatively short \tdelay\ period leads to an earlier onset of SF and higher SFRDs than models without such features. This results in earlier and deeper \Tb\ absorption trough minima in our simulations, as it sees earlier and more intense PopII SF than those of previous works. In general, we find that differences in \SFRDII\ largely translate to differences in \Ja\ and therefore \Tb, especially at lower redshifts and on smaller distance scales, while the \SFRDIII\ largely dominates higher redshifts and larger spatial scales (e.g. Figures \ref{fig:Power_Spectra}, \ref{fig:k-modes}). Overall, previously published 21-cm predictions tend to result in far lower or delayed SF, leading to a shallower and (mostly) later trough minimum than in our fiducial model.

As for the 21-cm power spectrum, we broadly find that power on scales $k \gtrsim 0.2\ \mathrm{Mpc^{-1}}$ at $z = 15$ mirrors the relative positions of \Tb($z=15$), and that power on all scales is (loosely) inversely correlated with \Tb\ over time (Figures \ref{fig:All_21cm} and \ref{fig:Power_Spectra}). Further, even though X-ray heating begins to wash out large-scale fluctuations at low $z$, we find that the 21-cm power spectra may be detectable with HERA starting at $z\simeq25$ for our fiducial model, beginning with $k\simeq 0.2\ \mathrm{Mpc^{-1}}$ and growing to larger $k$-modes with time (Figure \ref{fig:k-modes}).

Although the subtle differences in power between models makes distinguishing their properties via observation difficult, we found that changes to \tdelay\ result in the most dramatic differences in \Tb. Indeed, power spectrum upper limits (or detections) at $z>20$ from 1080\,hr of HERA observations under moderate foreground assumptions will begin to place significant constraints on \tdelay. The explicit recovery period for an individual halo following PopIII SF may only be obtained from simulations which include DM halo merger histories, such as the one presented here. 

We also compare our \Tb\ predictions with a model based on the 21cmFAST simulator \citep{Munoz22}, finding that this model yields far lower PopII SFRDs, leading to a much later and shallower absorption trough minimum than any of our models. The lack of stochastic DM halo merger trees and its relatively simplified SF and feedback prescriptions delays significant SF, yielding weaker and more uniform SF at high redshifts, which manifests as lower \PS\ than the models of this work. This once again spotlights the importance of including DM halo merger histories and SF prescriptions calibrated to state-of-the-art hydrodynamic simulations within theoretical models of CD.

In the future, we intend to improve the accuracy of our simulation framework by including more realistic prescriptions for \tdelay\ and the SF efficiencies which may vary by halo. We also intend to explore more alternative simulation assumptions/techniques with this framework, such as nonstandard DM scenarios. Improvements to simulation efficiency will further allow for more realizations and a more thorough exploration of the largely unconstrained parameter space of the first stars and CD. 

\section*{Acknowledgements}
We acknowledge support from NSF grant AST-2009309 and NASA ATP grant 80NSSC22K0629. All numerical calculations were carried out at the Ohio Supercomputer Center (OSC).

\section*{Data Availability}
The data underlying this article will be shared on reasonable request to the corresponding author.

\bibliography{main}{}
\bibliographystyle{aasjournal}

\bsp	
\label{lastpage}
\end{document}